\documentclass[letterpaper,11pt]{article}
\pdfoutput=1 

\usepackage{jheppub} 

\usepackage{float}   
\usepackage{subfig}

\usepackage[T1]{fontenc} 
\usepackage{tikz}
\usepackage{dsfont}
\usepackage{amsmath}
\usepackage{hyperref}
\def\ba{\begin{align}}\def\ea{\end{align}}
\def\beq{\begin{eqnarray}}\def\eeq{\end{eqnarray}}
\def\be{\begin{equation}}\def\ee{\end{equation}}
\def\ben{\begin{equation}}
\def\een{\end{equation}}
\def\bea{\begin{eqnarray}}
\def\eea{\end{eqnarray}}

\def\dn{{\rm{dn}}}
\def\cn{{\rm{cn}}}
\def\sn{{\rm{sn}}}
\def\bE{{\bf{E}}}
\def\bK{{\bf{K}}}
\def\bF{{\bf{F}}}
\def\tK{\tilde{K}}
\def\bOmega{\bar{\Omega}}
\def\bJ{\bar{J}}

\title{Fate of "Space-like singularities" in $c=1$ Matrix Model}
\begin{document}
\author[1]{Sumit R. Das,}
\author[2]{Shaun D. Hampton,}
\author[3]{Sinong Liu,}
\author[4]{Gautam Mandal.}

\affiliation[1]{Department of Physics and Astronomy, University of Kentucky, Lexington, KY 40506, U.S.A.}
\affiliation[2]{School of Physics,
Korea Institute for Advanced Study,
Seoul 02455, S. Korea}
\affiliation[3]{Yau Mathematical Sciences Center, Tsinghua University, Haidian District, Beijing 100084, China}
\affiliation[4]{International Center for Theoretical Physics, Shivakote, Bangalore 560089, India. }
\emailAdd{das@pa.uky.edu}\emailAdd{sdh2023@kias.re.kr}\emailAdd{sinongliu@mail.tsinghua.edu.cn}
\emailAdd{gautam.mandal@icts.res.in}

\abstract{A class of time dependent backgrounds in two dimensional String Theory leads to superluminal Liouville walls on the worldsheet. In the dual double scaled $c=1$ matrix model these backgrounds involve eigenvalues leaking out to infinity, and the collective field fluctuations become strongly coupled along space-like regions, resembling singularities. We realize these backgrounds as results of quantum quenches in the matrix model, retaining non-linear terms in the matrix potential, thus departing from a double scaling limit. {\color{black} Working in the fermion picture in a Thomas-Fermi approximation, we show that while the early time behavior of the phase space density near the maximum of the potential agrees with that obtained in the double scaled theory, at times of the order $(\log~N)$ the effect of the IR wall becomes significant. At later times, with a characteristic winding time of order $(\log N)^2$, folds on the fermi surface proliferate and eventually cover the allowed region in phase space densely. }Using action-angle variables, we show that the phase space density oscillates around a time independent {\color{black}and angle independent value} rapidly at late times. A coarse-grained density {\color{black} in the angle space relaxes to a time independent equilibrium value as a power law with an exponent largely independent of the details of the initial state. }Thus, the appearance of a space-like singularity is an artifact of the strict double scaling limit. We comment on the interpretation of the final state in String Theory.}


\begin{flushright}
\end{flushright}

\maketitle
\flushbottom

\section{Introduction and Summary}

It is notoriously difficult to discuss time dependent backgrounds in String Theory. Two dimensional non-critical String Theory \cite{2d,newhat} provides a rare example where this can be done to some extent, both on the worldsheet and in the dual {\em double scaled} $c=1$ gauged Matrix Quantum Mechanics of a single $N \times N$ Hermitian matrix, which reduces to a theory of free fermions in an inverted harmonic oscillator potential. More recently this gauge gravity duality has been understood at the non-perturbative level as well \cite{nonperturb,nonperturb2}.
Early discussions of such time dependent backgrounds appear in \cite{akk,alexandrov,strom,karc1,karc2,karc3,ddlm,dk}. These backgrounds have been mostly investigated using the matrix model where they appear as excited states represented by time dependent fermi surfaces. In this description the $1+1$ dimensional target space of the String Theory is emergent: the nature of this space-time follows from the behavior of fluctuations of the density of fermions (the collective field). Special cases were studied analytically. In one case the space-time had a moving mirror and scattering from the mirror leads to particle production \cite{alexandrov,strom,karc1,karc2,karc3,ddlm}. In another case, the space-time contains a space-like boundary where the self couplings diverge, akin to a space-like singularity \cite{dk,dsantos}. Other approaches to time dependent backgrounds in this theory include \cite{taka,timedepnew}.

More recently, \cite{bck} investigated these backgrounds from the point of view of the worldsheet. They found that for one class of backgrounds the Liouville wall remains time-like. In that case one can calculate the S-Matrix for scattering from this wall -- the result shows, among other things, particle production. However there is another class in which the wall turns superluminal. Now there is no asymptotic future infinity, and there is no S-Matrix. 

In \cite{dhl5} the emergent space-time which follows from the double scaled matrix model was examined in detail. It was found that for backgrounds of the first class, the space-time is regular everywhere apart from a time like region of large coupling which represents the Liouville wall. We will call this class the "moving mirror" backgrounds, and these are generalizations of the special cases studied in \cite{strom,karc1,karc2,karc3,ddlm}. Backgrounds of the second class (which lead to  superluminal Liouville walls) have emergent space-times with space-like regions of large coupling: these appear as space-like singularities of the semi-classical theory. However, the corresponding time dependent Fermi surfaces involve fermions which leak out to infinity -- which is possible in the double scaling limit, since the potential in which the fermions move is an inverted oscillator potential. This latter class is called "draining fermi sea": the backgrounds studied in \cite{dk,dsantos} are special cases. To know the true late time behavior of the model one needs to include an IR wall which keeps the number of fermions finite. However, inclusion of such an IR wall is a departure from the double scaled limit and  a finite $g_sN$ effect. These subleading effects are therefore crucial in understanding the true late time fate of the system.

In this paper we will re-examine the problem in the $c=1$ model with a matrix potential which keeps the fermions in a well. The specific potential we use is a {\color{black} double well potential}. In the double scaling limit the quartic term in the potential is then $O(1/g_sN)$. We will work in a limit where $N \gg 1$ and $g_s \ll 1$ so that $(g_s N)$ is fixed so that a Thomas Fermi approximation is still valid.
To track the dynamics in a well-defined manner we realize the solutions described in \cite{bck,dhl5} as resulting from an {\em abrupt quantum quench} where the system is abruptly quenched from the ground state of a deformed Hamiltonian to this Hamiltonian at some time. Quantum quench in matrix models were first studied in \cite{mandalmorita} in unitary matrix quantum mechanics where a coupling in the potential changes abruptly, leading to a rich landscape of dynamical transitions. Coupling quench in the $c=1$ double scaled model has been studied in \cite{dhl2} where the quench can be smooth or abrupt, leading to the conclusion that the late time emergent space-time is not generically smooth, except in finely tuned situations. Quenches in higher dimensional free fermion systems were studied in \cite{mandalkulkarni} for situations of direct relevance to cold atom physics, where it was shown that generically the system approaches a Generalized Gibbs Ensemble (GGE).

In this work we use, following \cite{mandalkulkarni}, the Liouville equation for the phase space density to study the late time dynamics of initial excited states which would be the dual descriptions of the String Theory backgrounds in the double scaling limit. {\color{black} We find that, as expected, while the presence of the quartic term does not change the early time behavior in the region far from the IR wall, at times of order $\log N$ fermions originally near the region near the maximum of the potential get reflected from the IR wall. 
This then leads to the formation of folds on the Fermi surface, with a characteristic winding time of order $(\log N)^2$. As time proceeds, multiple folds proliferate and the entire region between the orbits of lowest and highest energies gets filled by filaments of alternating filled and empty regions, as in the examples studied in \cite{mandalkulkarni}. Interestingly, this happens regardless of whether the fermi surfaces in the corresponding double scaled theory are of the moving mirror class or the draining fermi sea class.

It has been known for a while that formation of folds on a fermi surface in phase space (in the Thomas-Fermi approximation) is a signature of large {\em quantum fluctuations} of the collective field which survive the classical Thomas-Fermi limit \cite{fold1, fold2}. This means that a {\em classical bosonization} in terms of a single classical collective field and its canonically conjugate momentum fails -- one needs to introduce an infinite set of additional fields in an effective classical description. A {\em quantum bosonization} should still hold.  In our case, at intermediate times the profile of the phase space density is complicated. However, in the long time limit there are essentially an infinite number of folds between the lowest and highest energy orbit. 

To understand this late time behavior it is useful to use action-angle variables as coordinates in phase space. In terms of these variables, this problem can be then mapped to the problem of free fermions without any external potential living on a circle -- the time evolution of the phase space density for the latter problem has been analyzed in \cite{mandalkulkarni}, whose ideas we use. The result is that at late times the phase space density $u(J,\theta,t)$ oscillates around the time independent distribution, which is simply the number of fermions with a given energy, which is of course conserved. This then represents a Generalized Gibbs Ensemble. Averaging over an energy (or action) {\color{black} with some weight function leads to a} coarse-grained phase space density which asymptotes to this equilibrium distribution as a power law in time, while the exact distribution never becomes stationary. The exponent is universal, in the sense that it depends only on the behavior of the initial phase space density near the points of minimum and maximum energies.

In this paper, we have used a particular IR completion of the double scaled potential. The details of our results are not of course universal, and depend on the particular IR completion. However the main conclusion, i.e. that the phase space density approaches a time independent constant value should be valid for all IR completions. In this sense the late time fate of the space-like singularity is robust.

The meaning of this late time state in String Theory is less clear. Our work shows that a departure from the double scaling limit is necessary to understand the late time fate of these time dependent backgrounds. The double scaling limit corresponds to a critical point where the 't Hooft-Feynman diagrams form a continuous surface which is identified with the world-sheet of the dual String Theory. Away from this limit the world sheet remains discrete, and it is not clear how to formulate the String theory. A more detailed understanding would require understanding finite $N$ effects, beyond the Thomas-Fermi limit, perhaps along the lines of \cite{finiteNcollective,corley,jr1,rdmk,qoscillator,dharmandal,gautam,dmw,son}.
Interestingly, however, it has been shown that for certain $c=0$ matrix integrals it is nevertheless possible to define a continuous world sheet theory away from the double scaling limit \cite{gopa, mazenc}. It will be interesting to explore this possibility for matrix quantum mechanics.

In \autoref{secone} we review the essential features of the time dependent solutions in the double scaling limit and the nature of the emergent space-times. In \autoref{sectwo} we realize these time dependent solutions as results of a quantum quench in the presence of a quartic term in the potential (with the coefficient of the quartic term $\sim \frac{1}{Ng_s}$ relative to the quadratic term) and derive the equation for the fermi surface at any time.  In \autoref{secthree} we determine these fermi surfaces numerically and show that at late enough times they form folds and filaments filling the allowed region of the phase space. In \autoref{secfive} we introduce action angle variables in phase space, determine the phase space density in these coordinates and demonstrate that at late times a coarse grained density approaches a time independent equilibrium value for the given total energy. \autoref{secsix} is the concluding section. \autoref{doubscale} provides details of the double scaling limiting procedure. \autoref{maxmin} provides details of the evaluation of the time of fold formation. \autoref{action-angle} introduces the action-angle variables for the one-particle motion in the double-well potential. Building on this, \autoref{power-law-relaxation} derives the behaviour of the angle variables on the Fermi surface near points where the corresponding action is extremal and determines the relaxation exponent. In \autoref{secfour} we recap the results of \cite{fold1,fold2} relating the presence of folds with large quantum fluctuations of the collective field. \autoref{appboson} reviews standard bosonization formulae for chiral bosons.

\section{Time dependent Backgrounds in 2d String Theory}
\label{secone}

The worldsheet action for the class of backgrounds we will consider in this paper is given by \cite{akk,bck}
\ben
S_{WS} = \frac{1}{4\pi} \int d^2 \sigma \sqrt{h} \left( -(\nabla t)^2 + (\nabla \phi)^2 - 2 \phi R^{(2)}(h) - 4\pi \mu \phi e^{2 \phi }+e^{(2-r)\phi} e^{rt}\right)~,
\label{0-1}
\een
where $t (\sigma)$ is the target space time, $\phi(\sigma)$ is the Liouville field, $\mu$ is the worldsheet cosmological constant and $r$ is a real number. Worldsheet calculations with this action were performed in \cite{bck}. For $0 < r < 1 $ the Liouville wall, where the potential $V_{WS} = - 4\pi \mu \phi e^{2 \phi }+e^{(2-r)\phi} e^{rt} \sim 1$, moves at a sub-luminal speed which approaches $v \sim - r/(r-1)$ at late times. There is a well defined (target space) wall S-Matrix describing massless particles which includes, among other things, particle production. However for $1 < r < 2$ the Liouville wall becomes super-luminal after some time. Then there is no future asymptotic region and no obvious S-matrix. 

These backgrounds are represented by excited states in the singlet sector double scaled dual matrix model \cite{2d} . In the following we will use the conventions of \cite{polclass}. (See \autoref{doubscale} for a review of double scaling.) The action of the matrix model is 
\ben
S = \beta \int dt ~{\rm Tr} \left[ (D_t M)^2 - U(M) \right]~, \qquad \beta = \frac{N}{\kappa}
\label{0-2}
\een
and $\kappa$ is a coupling. In this paper we will consider a polynomial of the form
\ben
U(M) = -\frac{1}{2}M^2 + \frac{1}{4} M^4 + \cdots~.
\label{0-3}
\een
The double scaling limit is $N \rightarrow \infty$ and the fermi energy $-\mu_F \rightarrow 0$ with \footnote{Our conventions follow \cite{polclass}.}
\ben
g_s =- \frac{1}{2 \beta \mu_F} 
\label{0-4}
\een
and $\kappa$ held fixed. In this limit the theory reduces to a theory of $N$ free fermions with a second quantized Hamiltonian
\ben
H_{ds} =\frac{1}{g_s} \int dx \left[ \frac{g_s^2}{2} (\partial_x \psi^\dagger)
(\partial_x \psi) - \frac{1}{2} x^2 \psi^\dagger \psi \right]~, \qquad \int dx ~\psi^\dagger \psi = N~,
\label{0-6}
\een
where $x$ is defined in terms of the eigenvalue $\lambda$ by (\ref{0-5a}).
The connection to two dimensional string theory can be best understood in  terms of a collective field $\rho (\lambda)$ and its canonically conjugate momentum $\Pi_\rho (\lambda)$. At the operator level these are given by \cite{jevsak, dasjev}
\ben
\rho (\lambda) = \frac{1}{\beta} \Psi^\dagger \Psi =  \frac{1}{\beta} \delta (\lambda \cdot I - M) ~, \qquad \Pi_\rho = -i \frac{\delta}{\delta \rho}~.
\label{0-7}
\een
Double scaling is now achieved by definining a scalar $\varphi(x)$ and its canonical conjugate $\Pi_\varphi$ defined in (\ref{0-8a}), leading to the 
double scaled Hamiltonian 
\ben
H = \int dx \left[ \frac{g_s^2}{2} \Pi_\varphi (\partial_x \varphi) \Pi_\varphi + \frac{\pi^2}{6g_s^2} (\partial_x \varphi)^3- \frac{x^2}{2g_s^2} (\partial_x\varphi)  + \frac{1}{2g_s^2} (\partial_x\varphi) \right]~.
\label{0-9}
\een
The last term comes from a Lagrange multiplier which imposes the condition $\int dx \psi^\dagger \psi = N$. Note that we are using rescaled variables: in terms of these the Fermi level is $-1/2$ \cite{polclass}.

As is well known, fluctuations around the ground state classical solution of the Hamiltonian (\ref{0-9}) are described by a relativistic massless scalar field moving in $1+1$ dimensions $(x,t)$ with space dependent self coupling which diverges at $x=1$. It is usually convenient to make a coordinate transformation to go to new coordinates $(q,t)$ in which the background metric is conformal to Minkowski space-time. The self coupling then diverges at $q=0$. At the perturbative level this is mirror which reflects small fluctuations, and partitions two halves of Minkowski space. These collective field fluctuations are directly related to the "massless tachyon" of two dimensional string theory \cite{dasjev}. These are actually related by a spatial transform which is nonlocal at the string scale. This transform encodes gravitational interactions \cite{polnat}. However for the questions addressed in this paper we are not interested in sub-string scale physics so that this transform will not play a role.

In either formulation the classical limit is then seen to be $g_s \rightarrow 0$. In this limit we can use the Thomas Fermi approximation \cite{BIPZ} in the fermionic field theory \cite{fermionic}. The expectation value of the phase space density of fermions in some state $|\chi \rangle$
\ben
u (x,p,t) = \langle \chi | \left( \int dy ~\psi^\dagger (x-\frac{y}{2})\psi (x+\frac{y}{2})e^{ipy/g_s}\right) | \chi \rangle
\label{0-7a}
\een
is then either zero or one. The filled region is bounded by a fermi surface.
This quantity $u(x,p,t)$ then obeys the Liouville equation 
\ben
\left[ \partial_t -U^\prime (x) \partial_p + p\partial_x \right] u(x,p,t) = 0~.
\label{0-8}
\een
In the bosonic description this classical limit corresponds to a classical solution $\varphi_0 (x,t),( \Pi_\varphi)_0 (x,t)$ which are given by \cite{polclass}
\ben
\partial_x \varphi (x,t) = g_s\int \frac{dp}{2\pi g_s}~u(x,p,t)~, \qquad 
(\partial_x\varphi)\Pi_\varphi (x,t) = -\int \frac{dp}{2\pi g_s^2}~p~u(x,p,t)~.
\label{0-8a}
\een

The ground state in the Thomas Fermi limit  is given by
\ben
u_0 (x,p) = \Theta (x^2-p^2-1)~, \qquad \partial_x \varphi_0 = \frac{1}{\pi}\sqrt{x^2-1}~.
\label{0-9a}
\een
Small deformations of the fermi surface then correspond to fluctuations of the collective field around the ground state solution. 

The excited state of the double scaled theory with $U(x) = -x^2/2 $ dual to the string theory background (\ref{0-1}) is then given by a time dependent solution  
\ben
u_0(x,p,t) = \theta \left( x^2 - p^2 - e^{rt} (x-p)^r - 1 \right)~,
\label{0-10}
\een
where we have rescaled quantities to set the rescaled Fermi energy to be unity. The argument in the Heaviside theta function in (\ref{0-10}) is a {\em time dependent fermi surface}. Each point on the fermi surface moves along a phase space trajectory which, for an inverted oscillator potential is given by
\ben
x(t)\pm p(t) = e^{\pm t}( x(0)\pm p(0) )~.
\label{0-11}
\een
Each point on the initial fermi surface $ x^2 - p^2 - (x-p)^r = 1$ has a different energy, which is why the state is an excited state with non-trivial time dependence. The backgrounds considered in \cite{strom,karc1,karc2,karc3,ddlm} correspond to $r=1$, while those considered in \cite{dk,dsantos} have $r=2$. \footnote{  \cite{dk} also considered solutions where the sign of the time dependent term in (\ref{0-10}) is positive rather than negative. This is called an "opening hyperbola" solution in which fermions cross to the other side of the inverted oscillator potential. We will not consider this case in this paper.}

In \cite{dhl5} it was shown that the time dependent fermi surface of (\ref{0-10}) remains quadratic, i.e. a constant $x$ line intersects the fermi surface precisely twice at $P_\pm (x,t)$. The classical collective field and conjugate momentum can be then calculated using (\ref{0-8a}), e.g.
\ben
\partial_x \varphi (x,t) = \frac{1}{2\pi} (P_+(x,t) - P_-(x,t))~, \qquad (\partial_x\varphi) \Pi_\varphi = - \frac{1}{4\pi g_s^2} (P_+(x,t)^2 - P_-(x,t)^2)~.
\label{0-12}
\een
The fluctuations around this classical background are again described by a $1+1$ dimensional massless scalar field with space-time dependent couplings. Furthermore, using \cite{alexandrov} one can construct the Minkowskian coordinates $(q,\tau)$ such that the kinetic term for the fluctuations becomes simply $\int d\tau dq \frac{1}{2} [(\partial_\tau \eta)^2 - (\partial_q \eta)^2]$. This leads to a determination of the global properties of emergent space-times and their Penrose diagrams.

In \cite{dhl5} it was found that the two different regimes $0 \leq r \leq 1$ and $1 < r < 2$ give rise to globally different kinds of space-time. In the first regime,
$0 \leq r \leq 1$, the space-time is geodesically complete with a time-like region of strong coupling which partitions it. In particular there are both past and future asymptotic regions. However, in the second regime, $1 < r < 2$ the space-times have space-like boundaries where the couplings diverge. Near this singularity the semi-classical approximation in collective field theory no longer holds. This latter situation is akin to a space-like singularity where the effective field theory fails.

\autoref{ronetwo} shows the profile of the fermi surfaces at various times for $r = \frac{\pi}{2}$. The general behavior is similar for any $r$ in this range.  \autoref{spacetimepi2} shows constant $t$ slices in the $(\tau, q)$ plane for $r = \frac{\pi}{2}$ and the corresponding Penrose diagram.
\begin{figure}[!h]
\centering
\includegraphics[width=2.5in]{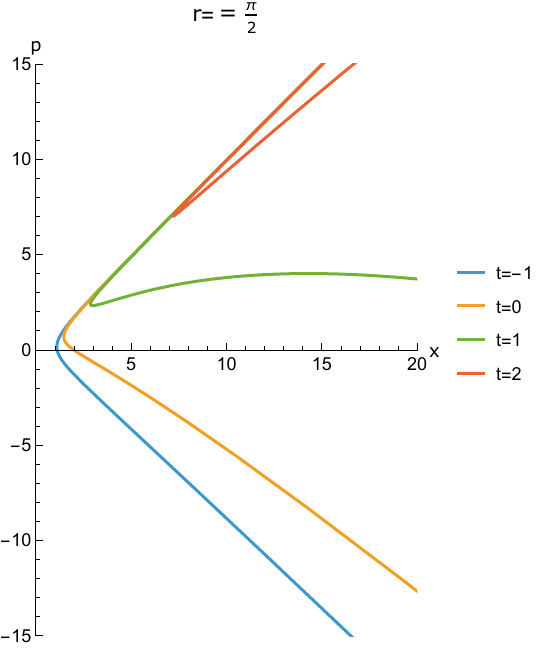}
\caption{Fermi surfaces at various times for $r = \frac{\pi}{2}$. This figure is taken from \cite{dhl5}.}
\centering
\label{ronetwo}
\end{figure}

\begin{figure}[!h]
\centering
\includegraphics[width=4.0in]{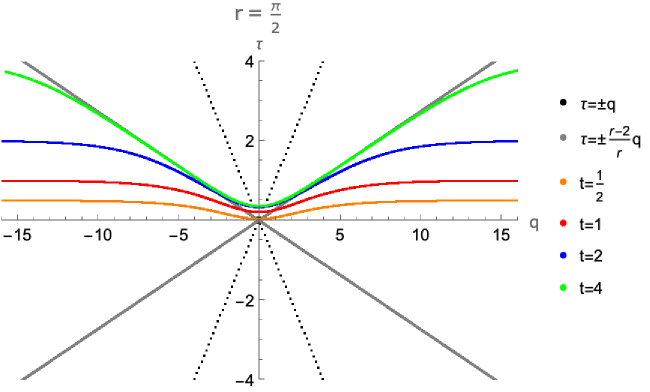}\\
\vspace{0.7cm}
\includegraphics[width=4.0in]{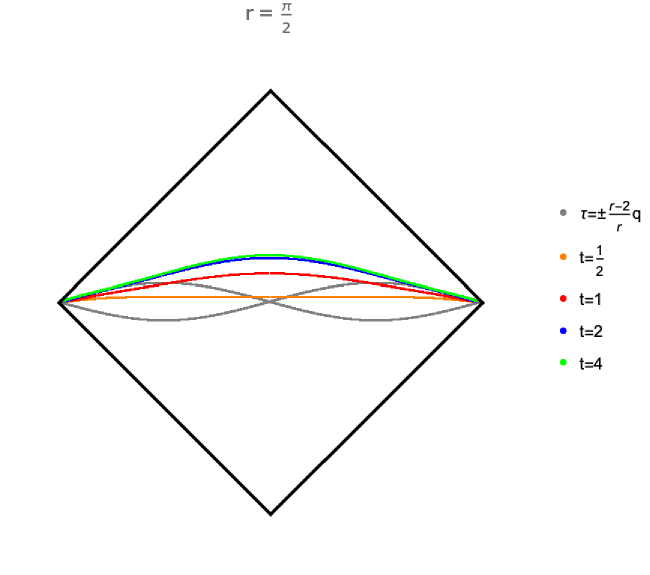}
\caption{Constant $t$ lines for $r = \frac{\pi}{2}$. Minkowski space (top) and Penrose diagram (bottom). This figure is taken from \cite{dhl5}.}
\centering
\label{spacetimepi2}
\end{figure}

The fermion theory is of course exact and should predict the time evolution. However, the fermi surfaces $x^2 - p^2 - e^{rt} (x-p)^r - 1$ in the double scaled theory lead to a leakage of the fermions to infinity as time evolves. This happens only because the inverted oscillator potential in the double scaling limit is unbounded, allowing particles to go away to infinity. To discuss the late time physics one needs to use a potential which is bounded, e.g. a $U(\lambda) = -\frac{1}{2} \lambda^2 + \frac{1}{4}\lambda^4$. The discussion of the double scaling limit, however, shows that an anharmonic term is suppressed by a power of $1/(\beta g_s) \sim 1/(g_s N)$. For a fixed $g_s$ this would mean a finite $N$. 

In the following sections we will study the late time behavior in a regime where $N$ is large but finite, but $g_s$ small with $g_s N$ finite, so that we can use a Thomas Fermi approximation. From (\ref{0-4}) we see that this corresponds to the fermi level of the original unscaled Hamiltonian below the quadratic maximum, which also means that the worldsheet provided by Feynman-'t Hooft diagrams is away from the continuum limit. By the same token the effect of the IR provided by an anharmonic term is not universal. Nevertheless, we expect that the late time behavior will be generic. A more complete treatment would require a treatment of finite $N$ effects, either in terms of suitable collective field operators \cite{finiteNcollective,corley,jr1,rdmk,qoscillator} or in terms of operator phase space density \cite{dmw,son} or other ideas of exact bosonization at finite $N$ \cite{gautam}.

\section{Time dependent Fermi surfaces from a quantum quench} 
\label{sectwo}

To discuss the time evolution properly we will realize these solutions as resulting from a quantum quench. We will work in a Thomas-Fermi approximation. A general solution of the Liouville equation (\ref{0-8}) with some initial data can be written as follows. Consider an initial profile at time $t = -T$
\ben
u(x,p,-T) =\Theta [ F(x,p) -1]~, \qquad \int \frac{dxdp}{2\pi g_s} u(x,p,-T) = N~,
\label{1-1}
\een
where $\Theta(z)$ is the Heaviside theta function.
 $F(x,p) = 1$ is then the initial fermi surface.
Consider a classical phase space trajectory $x(t),p(t)$ with the conditions
\ben
x(t)= x,~\qquad p(t) = p~.
\label{1-2}
\een
The point $(x,p)$ has evolved from a point $(x(-t-T),p(-t-T))$ at time $t = -T$. Since the phase space volume element is unchanged under time evolution we must therefore have
\ben
u (x,p,t) = u(x(-t-T),p(-t-T),-T) =\Theta [ F(x(-t-T),p(-t-T)) -1] ~.
\label{1-3}
\een
We can think of the initial fermi surface as the ground state of $N$ fermions with the single particle Hamiltonian given by
\ben
H_0 = -\frac{1}{2}F(x,p)~.
\label{1-4}
\een
Each fermion in the filled fermi sea subsequently moves according to the equations of motion which follow from the hamiltonian
\ben
H = \frac{1}{2} p^2 + U(x)~.
\label{1-5}
\een
Consider for example the double scaled theory $U(x) = -\frac{1}{2}x^2$ and a 
\ben
H_0 =-\frac{1}{2} F_{ds}(x,p) = \frac{1}{2} [p^2 -x^2 + e^{-rT}(x-p)^r ]~.
\label{1-6}
\een
Using (\ref{0-11}) we immediately see that $u(x,p,t)$ is given by (\ref{0-10}). This illustrates that this time dependent Fermi surface is a result of an abrupt quantum quench at $t=-T$.

We now are ready to investigate the effect of anharmonicity on the time evolution of an initial Fermi surface for an external potential 
\ben
U(x) = -\frac{1}{2}x^2 + \frac{g}{4} x^4~,\qquad g = \frac{1}{\beta g_s}~,
\label{1-7}
\een
where $x$ is the double scaled coordinate introduced in (\ref{0-5a}). The initial profile for the Fermi surface at $t=-T$ is chosen to be
\ben
F(x,p) = -p^2 +x^2 -\frac{g}{2}x^4 - e^{-rT}|x-p|^r ~.
\label{1-8}
\een
For small $x$ this profile agrees with the initial profile $F_{ds}(x,p)$ we considered in the double scaled theory, (\ref{1-6}). The quartic term ensures that the integral of the corresponding phase space density is finite.

The fermions evolve in time according to the equations of motion which follow from the Hamiltonian (\ref{1-5}) with $U(x)$ given by (\ref{1-7}). The trajectories $x(t), p(t)$ can be expressed in terms of Jacobi Elliptic functions. A point $(x,p)$ at time $t$ then evolves from a point $(x_{-T}, p_{-T})$ at time $t=-T$ 
\ben
x_{-T} (x,p,t) = \frac{x ~\dn (\omega (t+T);k) - \frac{p}{\omega} ~\sn (\omega (t+T);k) \cn (\omega( t +T);k)}{1- (1- \frac{g}{2} \frac{x^2}{\omega^2})\sn^2(\omega (t+T);k)}~,
\label{1-9}
\een
where
\ben
\omega(x,p)= \left[\frac{1}{2} ( 1 + \sqrt{1-4gE(x,p)})\right]^{1/2}~,~k^2 = \frac{2\sqrt{1-4gE(x,p)}}{1+\sqrt{1-4gE(x,p)}}~
\label{1-10}
\een
and in (\ref{1-9})
\ben
-E (x,p)= \frac{1}{2} \left[p^2 - x^2 +\frac{g}{2} x^4 \right]~.
\label{1-11}
\een
The Fermi surface at some time $t$ is then given by 
\ben
-p_{-T} (x,p,t)^2 + x_{-T}(x,p,t)^2 -\frac{g}{2}x_{-T}(x,p,t)^4 - e^{-rT}|x_{-T}(x,p,t)-p_{-T}(x,p,t)|^r =1~.
\label{1-12}
\een
Note that different points on the intial fermi surface (\ref{1-8}) have different energies, 
\ben
E(x,p)_{initial} = -\frac{1}{2} \left[ 1+ e^{-rT}|x-p|^r \right]~.
\label{1-12a}
\een
{\color{black} The point with the minimum and maximum values of the energy have the largest and smallest value of $|x-p|$. At these points on the fermi surface, denoted by $(x_{0,\pm}, p_{0,\pm})$ the tangent is given by $|x-p|=$(constant), i.e. $\frac{dp}{dx} = 1$. From (\ref{1-8}) this yields
\ben
p_{0,\pm} = x_{0,\pm} - g x_{0,\pm}^3~.
\label{1-12a}
\een
This, together with the equation $F(x_{0,\pm},p_{0,\pm})=1$ determine $(x_{0,\pm}, p_{0,\pm})$ which can be used to find the maximum and minimum energies $E_{\max}, E_{\min}$.  In \autoref{maxmin} we estimate these for $g \ll 1$. In the regime of interest $r < 2$ we find $E_{\min} \sim O(1)$ while 
$E_{\max} \sim g^{-r/2}$. Since surfaces of constant energy foliate the phase space, it is then clear that the fermi surface at a later time has to lie between the maximum and minimum energy orbits.}

Clearly, at early times, i.e. small $(t+T)$ the Fermi surface agrees with the double scaled theory in the region of small $(x,p)$. The quartic potential provides an IR wall -- close to this wall the solution will deviate from the double scaled result for all times. At sufficiently large times a fermion on the initial fermi surface at small $x$ will have sufficient time to go to the IR wall and get reflected back, modifying the fermi surface at small $x$, thus deviating from the double scaled theory. The earliest time this would happen can be estimated by calculating the time period of oscillation of the point on the initial fermi surface with minimum energy. 
A particle on the initial surface with energy $-E$ ($E > 0$) has a time period  of oscillation given by
\ben
T (-E)= 2 \int_{c}^{a} \frac{dx}{\sqrt{-2E + x^2 - \frac{g}{2} x^4}}~,
\label{1-13}
\een
where $0<c<a$ are the two turning points. Their explicit expressions are given in \eqref{a_and_c}. 
The integral can be performed in terms of the complete elliptic integral of the first kind,
giving
\ben
T (-E) = 2 \omega^{-1} \bK (k) ~,
\label{1-15}
\een
where $\omega$ is the frequency parameter defined in \eqref{1-10} for the orbit with energy $-E$.
For $gE \ll 1$ this gives
\ben
T (-E) \sim -\log (gE)~.
\label{1-16}
\een
{\color{black} Since $E_{\max} \sim g^{-r/2}$ (for $r < 2$)  for small $g$ this time is $\sim \frac{2-r}{2} \log (g) \sim \log (g)$. 
Therefore for our case where $g = 1/(\beta g_s) \sim O(1/N)$ and $E \sim O(1)$ this time is $O(\log~N)$. 
}

\section{Time evolution of Fermi surfaces}
\label{secthree}

In this  section we will determine the Fermi surface at various times by plotting (\ref{1-12}) and discuss the late time behavior.

Figures \ref{rhalft7} - \ref{rhalft100} show representative fermi surfaces with $r=0.5$ and an initial time $T=-5$ at various times. The quartic coupling $g = \frac{1}{\beta g_s} = 0.001$. In these figures the green curve is the initial fermi surface and the dotted curves are the trajectories of the mimimum and maximum energy particle on the fermi surface.

\begin{figure}[h]
\centering
\includegraphics[width=2.6in]{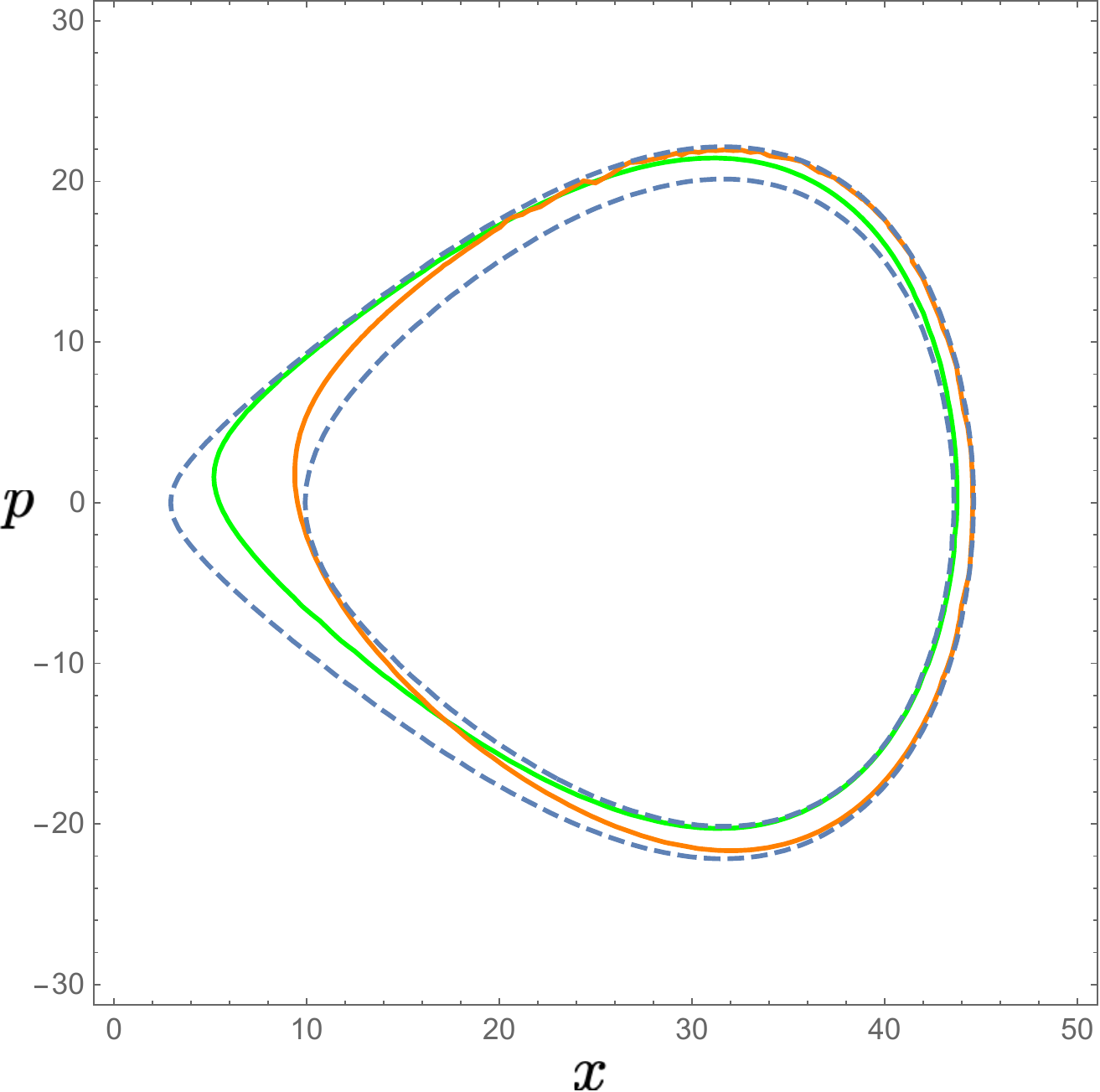}
\caption{ The green curve is the initial surface. Orange curve is Fermi surface at $t=7$ with $T=-5,r=0.5,g=0.001$. The dotted curves are maximum and minimum energy curves.}
\centering
\label{rhalft7}
\end{figure}

\begin{figure}[h]
\centering
\includegraphics[width=2.6in]{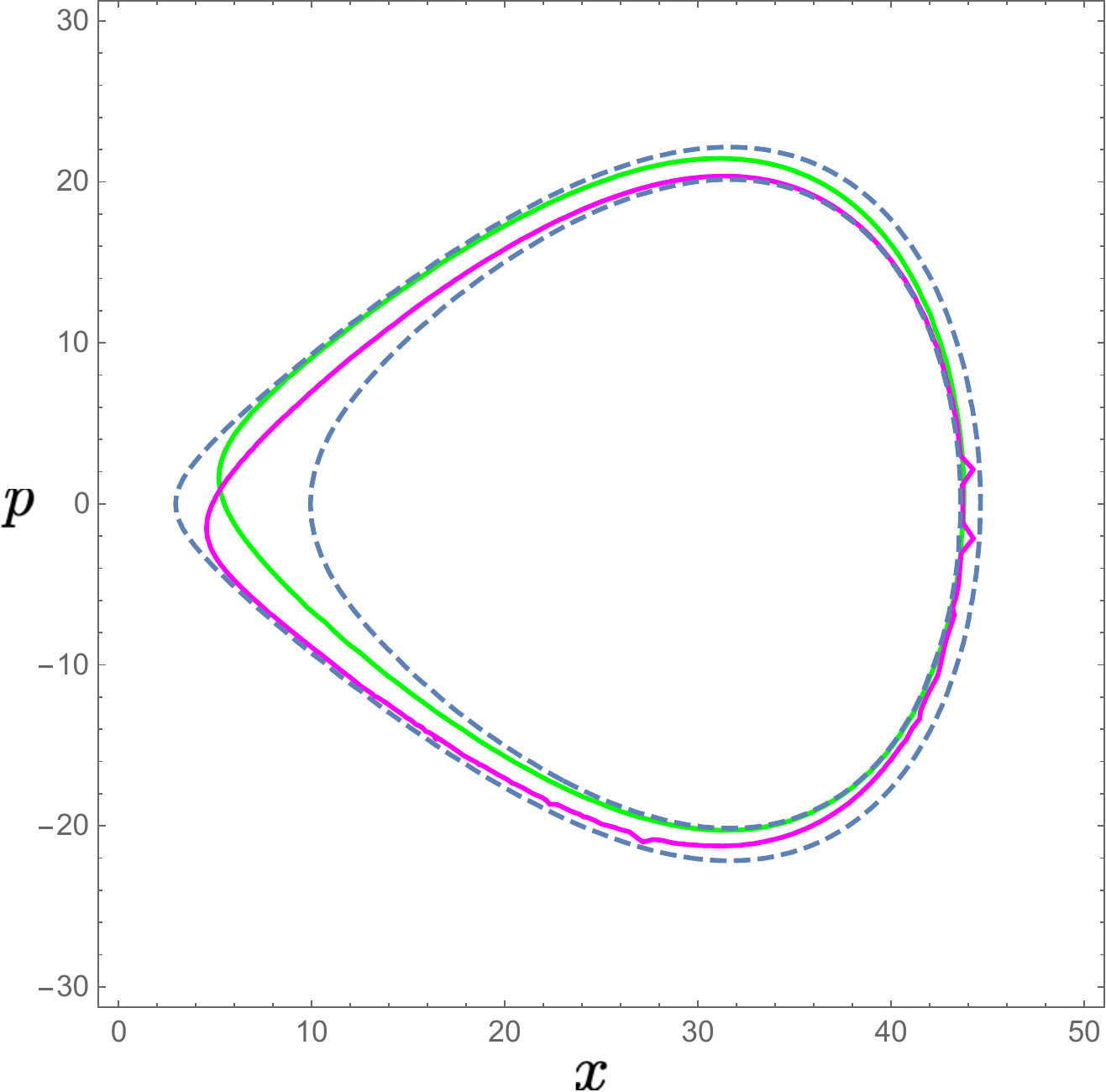}
\caption{The magenta curve is the Fermi surface at $t=10$ with $T=-5,r=0.5,g=0.001$. The green curve is the initial surface. The dotted curves are maximum and minimum energy curves.}
\centering
\label{rhalft10}
\end{figure}

Zooming on to the region of small $x$ in \autoref{rhalft7} it is clear that the motion of the fermi surface in this region at early times coincides with that in the double scaled theory. At an intermediate time $t=10$ the fermi surface still remains quadratic, as shown in \autoref{rhalft10}.

{\color{black} At sufficiently late times a fold appears on the surface, i.e. a given constant $x$ line intersects the surface more than twice. This is shown in  \autoref{rhalft20}. 
}

\begin{figure}[h]
\centering
\includegraphics[width=2.6in]{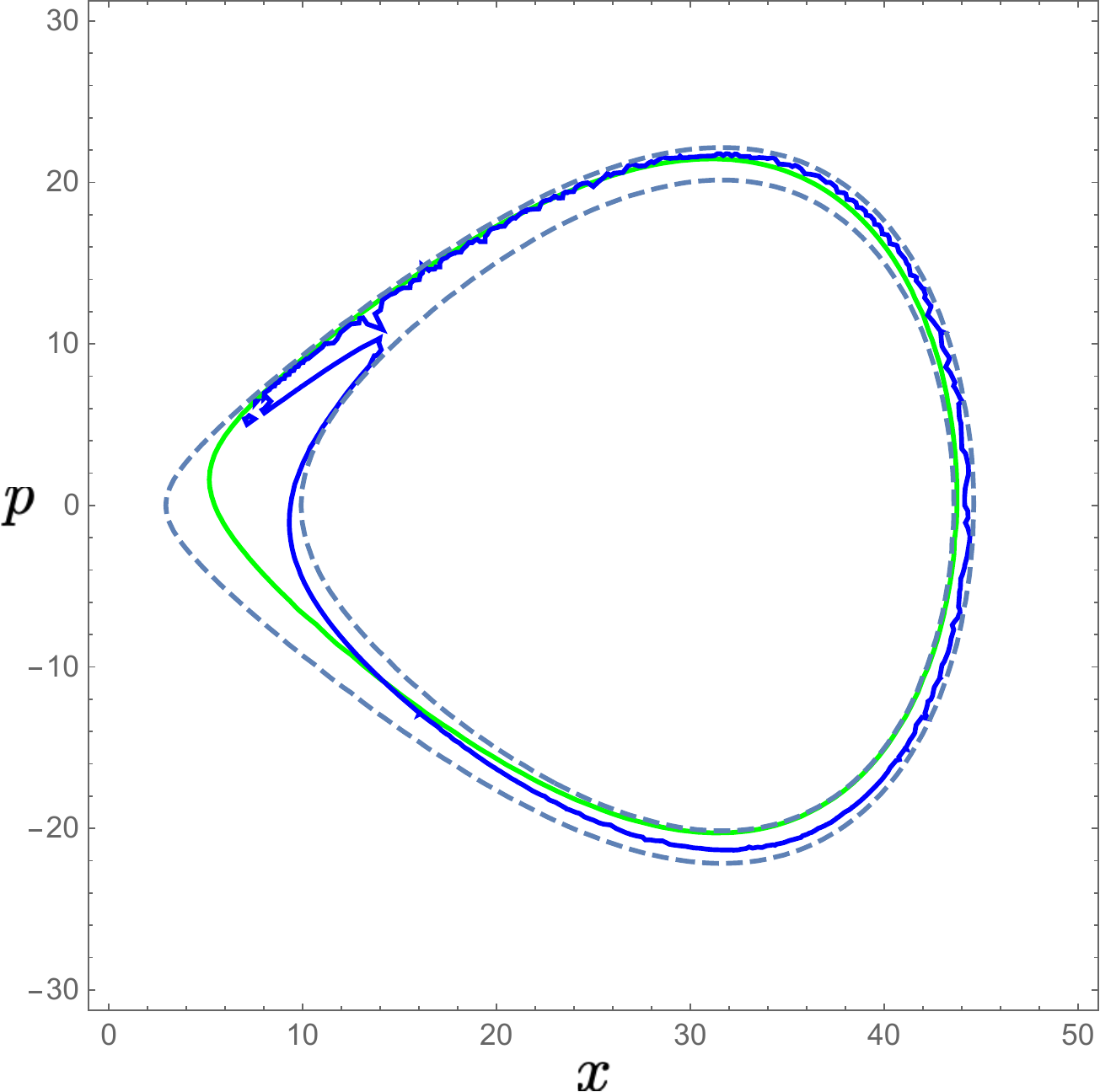}
\caption{The blue curve is the Fermi surface at $t=20$ with $T=-5,r=0.5,g=0.001$. The green curve is the initial surface. The dotted curves are maximum and minimum energy curves.}
\centering
\label{rhalft20}
\end{figure}

\begin{figure}[h]
\centering
\includegraphics[width=2.6in]{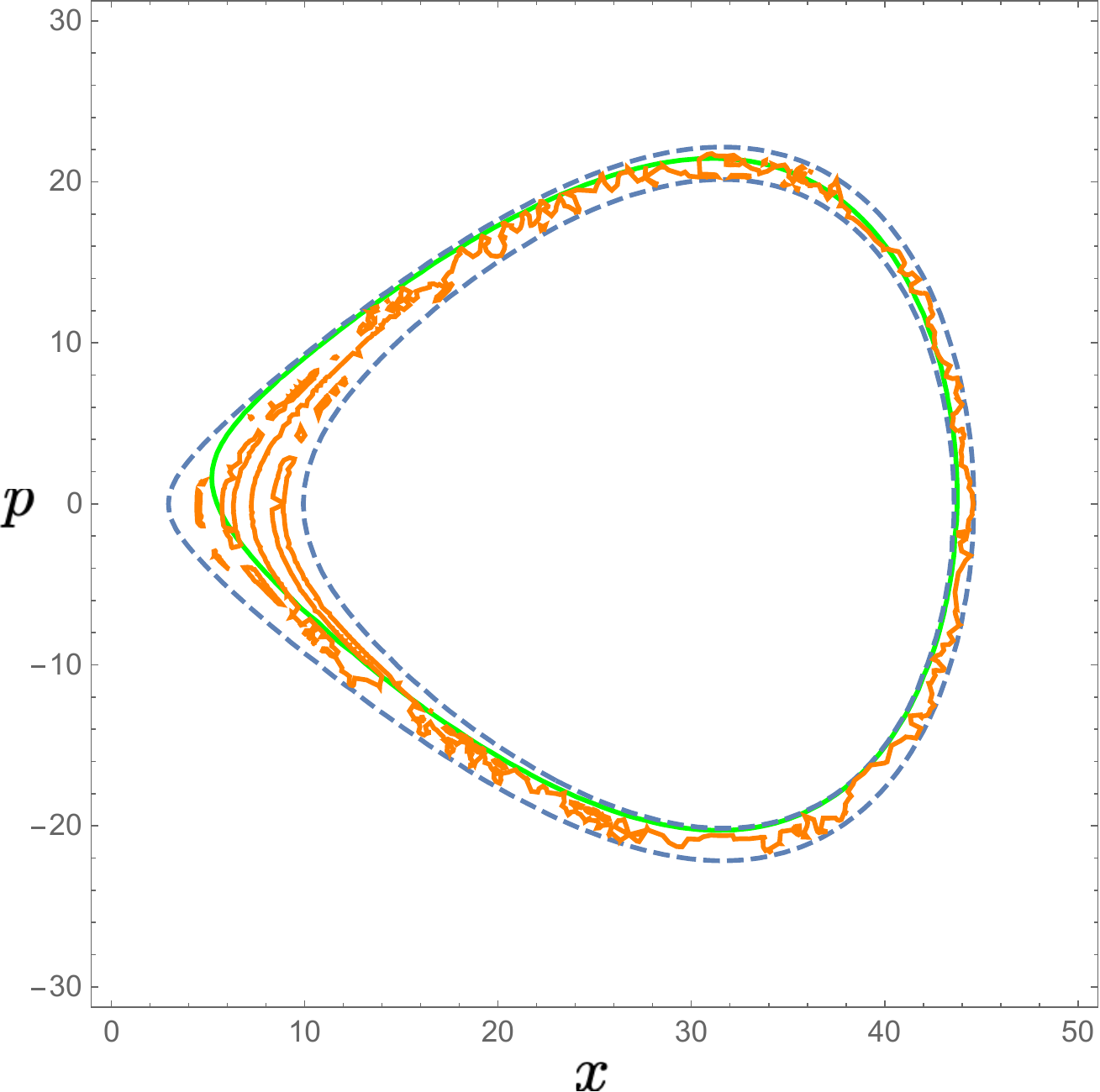}
\caption{The orange curve is the Fermi surface at $t=100$ with $T=-5,r=0.5,g=0.001$. The green curve is the initial surface. The dotted curves are maximum and minimum energy curves.}
\centering
\label{rhalft100}
\end{figure}

The growth of the folds is driven by the different angular velocities of orbits with different energies. As a result, the Fermi surface is continuously wound up by the accumulation of the relative angular displacement, leading to an increasing number of folds. A useful characteristic time is the winding time, defined as the time needed for the fastest orbit, with energy $-E_{\max}$, to get one period ahead of the slowest orbit, with energy $-E_{\min}$. This gives
\ben
T_{\rm winding}
=
\frac{2\pi}
{\frac{2\pi}{T(-E_{\max})}-\frac{2\pi}{T(-E_{\min})}}
\sim
(\log g)^2
\sim
(\log N)^2 .
\een
See \autoref{maxmin} for the estimate of $E_{\max}$, $E_{\min}$ and the winding time.

As time passes these folds proliferate and at very late times these folds form almost a ``space-filling'' curve in the entire region in phase space bounded by the maximum and minimum energy orbits. This is shown in \autoref{rhalft100}.

The Figures \ref{r32t31.5} - \ref{r32t150} show the fermi surfaces for $r = 1.5$. \autoref{r32t31.5} is for an early time $t=1.5$. This clearly shows that for such early times in the small $x$ region the fermi surface behaves like that in the double scaled theory.

\begin{figure}[h]
\centering
\includegraphics[width=2.6in]{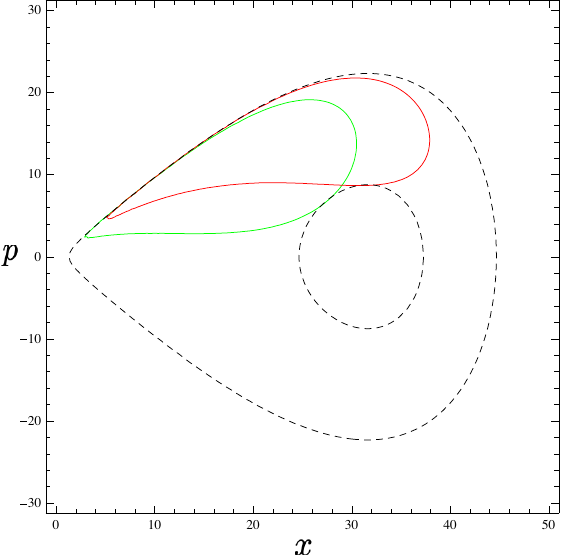}
\caption{The red curve is the Fermi surface at $t=1.5$ with $T=-1,r=1.5,g=0.001$. The green curve is the initial surface. The dotted curves are maximum and minimum energy curves.}
\centering
\label{r32t31.5}
\end{figure}

Departures from the double scaled behavior and formation of folds set in at intermediate times, as shown in Figures \ref{r32t36} and \ref{r32t310}.

\begin{figure}[h]
\centering
\includegraphics[width=2.6in]{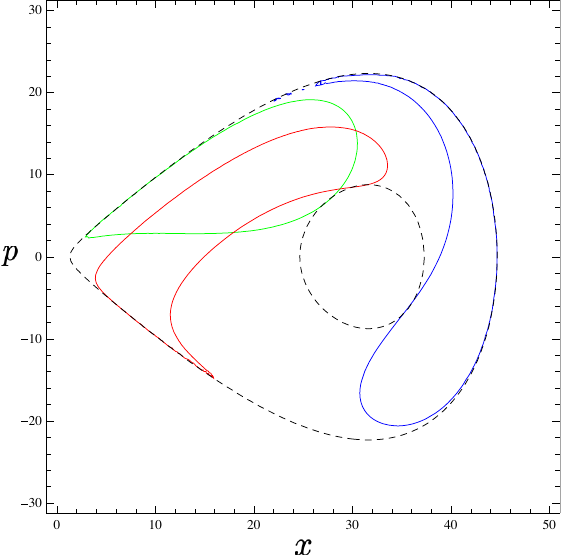}
\caption{The blue curve is the Fermi surface at $t=3$ with $T=-1,r=1.5,g=0.001$. The red curve is at $t=6$. The green curve is the initial surface. The dotted curves are maximum and minimum energy curves.}
\centering
\label{r32t36}
\end{figure}

\begin{figure}[h]
\centering
\includegraphics[width=2.6in]{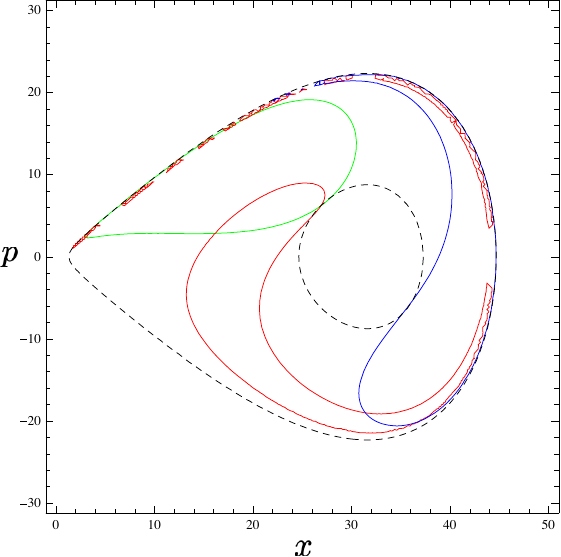}
\caption{The blue curve is the Fermi surface at $t=3$ with $T=-1,r=1.5,g=0.001$. The red curve is at $t=10$. The green curve is the initial surface. The dotted curves are maximum and minimum energy curves.}
\centering
\label{r32t310}
\end{figure}

Once again at very late times one tends to get a ``space-filling'' curve in the region between the minimum and maximum energy orbits, as shown in \autoref{r32t150}.

\begin{figure}[h]
\centering
\includegraphics[width=2.6in]{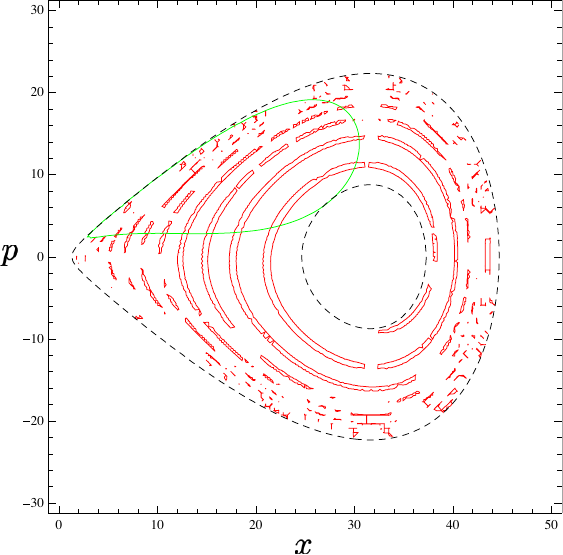}
\caption{The red curve is the Fermi surface at $t=150$ with $T=-1,r=1.5,g=0.001$. The green curve is the initial surface. The dotted curves are maximum and minimum energy curves.}
\centering
\label{r32t150}
\end{figure}

{\color{black} The initial profile used for $r=3/2$ in Figures \ref{r32t31.5} - \ref{r32t150} is qualitatively different from the $r=1/2$ in Figures \ref{rhalft7} - \ref{rhalft100}. In the former the initial Fermi surface does not wind around the center of the orbits, while in the latter it does. In the unwound case the filled region at late times is entirely contained in the region between the maximum and the minimum energy orbits of points on the Fermi surface, wheras in the wound case the filled region includes the interior. This is not a property of the value of $r$, but a choice of the initial time $T$. For $r > 1$ an initial profile which does wind around is shown in Figure \ref{r32T0t1.5} and \ref{r32T0t150}. In all cases the folds and filaments form between the maximum and minimum energy orbits.}

\begin{figure}[h]
\centering
\includegraphics[width=2.6in]{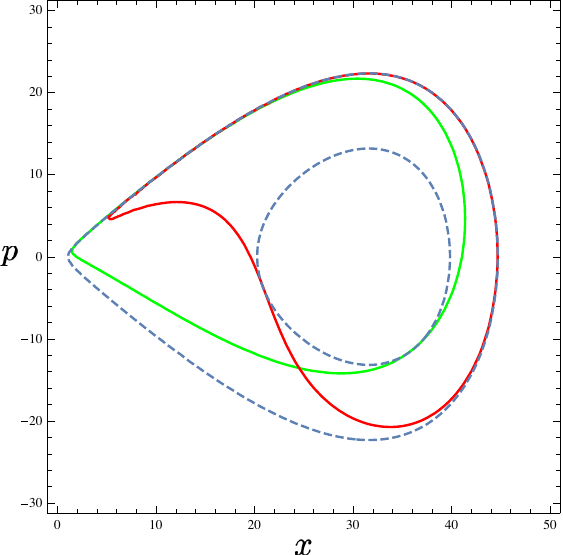}
\caption{The red curve is the Fermi surface at $t=1.5$ with $T=0,r=1.5,g=0.001$. The green curve is the initial surface. The dotted curves are maximum and minimum energy curves.}
\centering
\label{r32T0t1.5}
\end{figure}

\begin{figure}[h]
\centering
\includegraphics[width=2.6in]{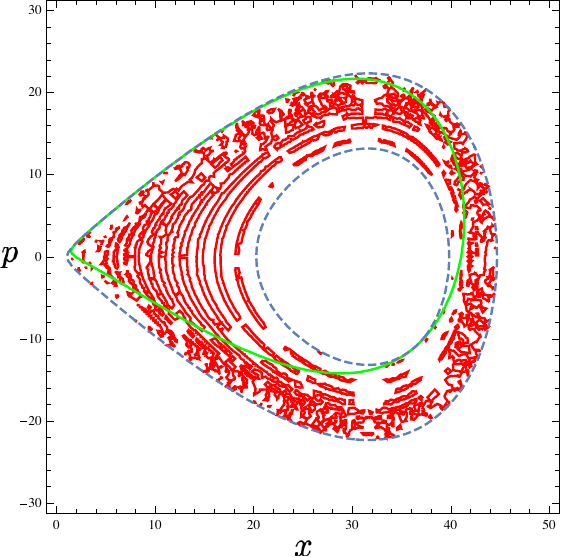}
\caption{The red curve is the Fermi surface at $t=150$ with $T=0,r=1.5,g=0.001$. The green curve is the initial surface. The dotted curves are maximum and minimum energy curves.}
\centering
\label{r32T0t150}
\end{figure}

\section{The late time fate}
\label{secfive}

The results of \autoref{secthree} show that away from the double scaling limit an initial fermi surface (\ref{1-8}) develops folds which proliferate. In fact folds are almost inevitable in a theory of non-relativistic fermions since fermions with higher momentum move faster. It was shown in \cite{fold1,fold2} that such folds are signatures of the presence of large quantum fluactations of the density operator. 
This means that the {\em classical bosonization formulae} (\ref{0-12}) do not work any more, even though a bosonization at the quantum level is possible. It is, however, still possible to describe the state in terms of a classical (Thomas-Fermi) phase space density. \autoref{secfour} contains a review of this argument. 

At late times, the phase space density develops thin filaments  which uniformly fill the region between the minimum and maximum energy orbits, pretty much like the systems discussed in \cite{mandalkulkarni}. Significantly, this outcome holds for both $r <1$ and $r > 1$. 

It is useful to discuss this final state in terms of action-angle variables in phase space. For our case these are given by
\bea
J(x,p) & = & \frac{1}{2\pi} \oint dx~p(x) = - \frac{1}{3\pi} \omega(x,p) \left[ \frac{2}{g} \bE (k (x,p)) -\frac{4}{g}\left(1-\omega(x,p)^2 \right) \bK (k (x,p)) \right]~, \nonumber \\
\theta (x,p) & = & 
\begin{cases}
\displaystyle
\frac{\pi}{\bK (k(x,p))} \bF (\phi(x,p), k(x,p))~, & p \cdot \operatorname{sgn} (x) > 0  \\[12pt]
\displaystyle
\frac{\pi}{\bK (k(x,p))} \left[ 2 \bK (k(x,p)) - \bF(\phi(x,p), k(x,p)) \right]~, & p \cdot \operatorname{sgn} (x) < 0
\end{cases}~.
\label{5-1}
\eea
Here $\bK (k), \bE (k)$ are the complete elliptic integrals of the first and second kinds, respectively and $\bF (\phi,k)$ is the incomplete elliptic function of the first kind with
\ben
\sin^2 \phi (x,p) = \frac{ 1- \frac{g}{2} \frac{x^2}{\omega^2 (x,p)}}{k^2}~.
\label{5-2}
\een
The functions $\omega(x,p), k(x,p)$ are defined in (\ref{1-10}). 
More details can be found in \autoref{action-angle}.
Since this is a canonical transformation the phase space density is
\ben
u (\theta, J, t) = u (x,p,t)~.
\label{5-3}
\een
With our convention that the physical orbit energy is $-E$, the trajectory of a point in phase space $(\theta,J)$ now obeys
\ben
\partial_t \theta = \Omega (J) = \frac{\partial (-E)}{\partial J}~, \qquad \partial_t J = 0~,
\label{5-4}
\een
thus mapping the problem to that of free fermions {\em without any external potential} on a circle. A trajectory in phase space is then characterized by a constant value of $J$ and
\ben
\theta (t) = \Omega (J) t + \theta_0~.
\label{5-5}
\een
The Liouville equation is
\ben
[ \partial_t + \Omega (J) \partial_\theta ] u (\theta,J,t) = 0~.
\label{5-6}
\een
In our case we have for a given point in phase space
\ben
 \Omega (J) =- \frac{\pi \omega}{\bK (k)}  = - |\Omega(J)| ~,
\label{5-7}
\een
where $k$ is the modulus defined in (\ref{1-10}).

The equations (\ref{5-5}) and (\ref{5-6}) then imply that given some initial phase space distribution $u_0(\theta,J)$ at time $t=0$, the phase space density at any given time is
\ben
u(\theta,J,t) = u_0( \theta (-t), J)=\theta [{\tilde{F}}(\theta,J)-1]~,
\label{5-8}
\een
where ${\tilde F}(\theta,J)=F(x,p)$ as defined in \autoref{sectwo}, where we have taken the initial time $T=0$. Due to the quadratic nature of the initial profile (\ref{1-6}) and the expression for the energy (\ref{1-12a}), a constant $J$ (which is also constant $E$) line intersects the initial fermi surface at most twice. When this intersects the fermi surface twice we denote these points by $\theta_{0,\pm} (J)$. In this case, all fermions with action $J$ (and therefore energy $E$) initially fill the entire angular interval $[\theta_{0,-} (J),\theta_{0,+} (J)]$. When there is only one intersection or no intersection, $\theta_{0,\pm} (J)$ should be understood through the occupied angular interval, with $\theta_{0,+} (J) - \theta_{0,-} (J) = 0~ \text{or} ~2\pi $, depending on the actual situation.

Then we can write, using (\ref{5-5}),
\ben
u (\theta,J,t) = \sum_{m=-\infty}^\infty \Theta \left[ (\theta_{0+}(J) - \left|\Omega (J)\right| t )-(\theta + 2\pi m ) \right] 
\Theta \left[ (\theta + 2\pi m) - (\theta_{0-}(J)- \left|\Omega (J)\right| t ) \right]~,
\label{5-9}
\een
where the sum over the integer $m$ is required to ensure periodicity of $u(\theta,J,t)$ under $\theta \sim \theta + 2\pi$. Using Poisson resummation
\ben
\sum_m f (2\pi m) = \sum_k \frac{1}{2\pi}\int dz~f(z)~e^{-ikz}
\een
and performing the $z$ integral using the explicit form of $u(\theta,J,t)$ we get, after separating the $k=0$ term,
\bea
u (\theta,J,t) & = & \frac{1}{2\pi} \left( \theta_{0+}(J) - \theta_{0-}(J) \right) \nonumber \\ + 
& & \sum_{k \neq 0} \frac{1}{2\pi ik} \left[ {\rm{exp}} \left(ik (\theta_{0+}(J)-|\Omega(J)|  t-\theta)\right) - {\rm{exp}} \left(ik (\theta_{0-}(J)- |\Omega(J)|  t-\theta) \right) \right] ~.\nonumber \\
\label{5-10}
\eea
The first time independent term is the total number of fermions with a given $J$, or given energy. Since the fermions are free, this is a conserved quantity.

{\color{black}
The  time dependent terms are oscillatory in time. At late times one expects that they vanish. To determine the way this relaxation occurs it is useful to define a coarse grained phase space density by averaging over $J$ with some kernel whose width is determined below. 

It is easier to peform this averaging by integrating over $\Omega$,  with a kernel $K_{\sigma} (\Omega)$
with a central value $\bOmega$ and a width $\sigma$:
\ben
u_{cg,\sigma} (\bJ,\theta,t) = \int d\Omega~ K_{\sigma} (\Omega) u (J (\Omega),\theta,t) = \int_{-\infty}^\infty d\beta~\tK_\sigma (\beta)\int_{\Omega_1}^{\Omega_2} d\Omega~e^{-i\beta \Omega} u (J(\Omega,\theta,t))~,
\label{5-12}
\een
where $\bar{J}$ is the value of $J$ at $\bOmega$. 
$\tK_\sigma (\beta)$ is the Fourier transform of $K_\sigma (\Omega)$ 
and $\Omega_1 = | \Omega (J(E_{\min})) |,~\Omega_2 = | \Omega (J(E_{\max})) |$ are the smallest and largest value of $\Omega$ on the initial Fermi surface, repectively. Clearly $\theta_{0, + } (\Omega_{1,2}) - \theta_{0, - } (\Omega_{1,2}) = 0 \pmod {2\pi}$.
In (\ref{5-12}) all quantities in the integrand should be expressed as functions of $\Omega$.

We are interested in evaluating (\ref{5-12}) at late times, much larger than the width $\frac{1}{\sigma}$ of the kernel $\tK_\sigma (\beta)$,
\ben
t \gg \frac{1}{\sigma}~.
\label{5-19}
\een
The integral over $\beta$ is then effectively over small values of $\beta$. Now, the integral over $\Omega$ can now be performed along the lines of \cite{mandalkulkarni}. Consider the contribution from a term with $k > 0$. 
We need to evaluate 
\ben
\int_{\Omega_1}^{\Omega_2} \frac{d\Omega}{2\pi i k} \left[ e^{[ik (\theta_{0,+} - \Omega t - \theta)-i\beta\Omega]} - e^{[ik (\theta_{0,-} - \Omega t - \theta)-i\beta\Omega]}\right]~.
\label{5-13}
\een
Now deform the contour of integration to the contour in the lower half plane: $[\Omega_1, \Omega_1 - i\Lambda, \Omega_2- i\Lambda, \Omega_2]$ with $\Lambda \gg 1$,
so that we get for the first term in (\ref{5-13})
\ben
\left[ \int_{\Omega_1}^{\Omega_1-i\Lambda}
-\int_{\Omega_2}^{\Omega_2-i\Lambda}+\int_{\Omega_1 - i\Lambda}^{\Omega_2-i\Lambda}\right]\frac{d\Omega}{2\pi i k}~
 e^{[ik (\theta_{0,+} - \Omega t - \theta)-i\beta\Omega]}~.
\label{5-14}
\een
The third term in (\ref{5-14}) decays exponentially for large $t$, provided $kt+\beta>0$. In the first and second terms, the dominant contributions come from the region near the real axis, where $\Omega = \Omega_1 - i \eta$ in the first term and $\Omega = \Omega_2 - i \zeta$ and taking the range of $(\eta,\zeta) \in (0,\infty)$. Therefore the leading effect can be estimated by looking at the expansion of $\theta_{0,+} (\Omega)$,
\bea
\theta_{0,+}(\Omega) & = & \theta (\Omega_1) + \gamma_1^+ (\Omega - \Omega_1)^{\alpha_1^+} + \cdots~, \qquad \Omega \sim \Omega_1~, \nonumber \\
\theta_{0,+}(\Omega) & = & \theta (\Omega_2) + \gamma_2^+ (\Omega_2 - \Omega)^{\alpha_2^+} + \cdots~, \qquad \Omega \sim \Omega_2~.
\label{5-15}
\eea
Expanding the integrand in small $\eta,\zeta$ and performing the integrals of $\eta,\zeta$ one gets for large $t$,
\ben
\int_{\Omega_1}^{\Omega_2} \frac{d\Omega}{2\pi i k} e^{[ik (\theta_{0,+} - \Omega t - \theta)-i\beta\Omega]} \sim O \left(\frac{1}{(\beta + kt)^{\alpha_1^++1}}\right) + 
O \left( \frac{1}{(\beta + kt)^{\alpha_2^++1}} \right)
\label{5-16}
\een
upto terms which oscillate as a function of time. For details of the integrals, see a completely analogous calculation in \cite{mandalkulkarni}.
The estimation of the second term in (\ref{5-13}) is entirely similar. Expanding 
\bea
\theta_{0,-}(\Omega) & = & \theta (\Omega_1) + \gamma_1^- (\Omega - \Omega_1)^{\alpha_1^-} + \cdots~, \qquad \Omega \sim \Omega_1  \nonumber \\
\theta_{0,-}(\Omega) & = & \theta (\Omega_2) + \gamma_2^- (\Omega_2 - \Omega)^{\alpha_2^-} + \cdots~, \qquad \Omega \sim \Omega_2
\label{5-17}
\eea
and one gets estimates like (\ref{5-16}) with $\alpha_{1,2}^+ \rightarrow \alpha_{1,2}^-$. Therefore the time dependent parts of $u_{cg}$ behave as
\ben
 \int d\beta~ \tK_\sigma (\beta) \frac{1}{(\beta + kt)^{\alpha +1}}~, \qquad
\alpha = {\rm min} (\alpha_{1,2}^{\pm})~.
\label{5-18}
\een
For late times (\ref{5-19}) we need to consider only values of $\beta$ such that $ t \gg \beta$. Therefore
we conclude that the coarse grained density approaches a time independent steady state value with a power law decay $t^{-(\alpha +1)}$. For a coarse graining scale $\sigma$ which is small compared to the scale of the variation of $\theta_{0,\pm}(\Omega)$, the steady state value is given by
\ben
u_{cg,\sigma} (\bJ, \theta) \sim \frac{1}{2\pi} \left( \theta_{0+}(\bOmega) - \theta_{0-} (\bOmega) \right)~.
\label{5-20}
\een
This is of course independent of $\theta$, a signature that the system asymptotes to a Generalized Gibbs Ensemble characterized by the conserved number of fermions at each value of the energy. Note that \eqref{5-20} represents this conserved number.

Unlike the true $u(J,\theta,t)$ the coarse grained quantity 
$u_{cg,\sigma}$ is not $0$ or $1$. Let us denote the action for the maximum and minimum energy orbits by $\bJ_{\max,\min}$.
For fermi surface profiles shown in Figures \ref{rhalft7} - \ref{rhalft100}, $u_{cg,\sigma} (\bJ, \theta) = 1 $ for $\bJ < \bJ_{\min}$ and $u_{cg,\sigma} (\bJ, \theta) = 0$ for $\bJ > \bJ_{\max}$ and interpolates between $1$ and $0$ in-between. For initial fermi surfaces which are of the type shown in Figures \ref{r32t31.5} - \ref{r32t150}, $u_{cg,\sigma} < 1$ for $\bJ_{\min} < \bJ < \bJ_{\max}$ and vanishes otherwise.


It is significant that the exponent $\alpha$ of the relaxation to this steady state value does not depend on the details of the initial profile -- this only depends on how the $\theta_{0,\pm}$ behave near their extrema.
In \autoref{power-law-relaxation}, expansion of the intersections of the constant-energy orbit with the initial Fermi surface around the extremal energies $E_{\min, \max}$ shows that their splitting is proportional to $|E-E_{\min, \max}|^{1/2}$. This leads to the leading behavior of
$\theta_{0,\pm} (E) \sim |\Omega-\Omega_{1,2}|^{1/2}$. The explicit energy dependence of $k$, $\omega$, and $\bK(k)$ appearing in \eqref{5-1} contributes only at higher orders. As a result, we obtain
\begin{equation}
\alpha_1^\pm=\alpha_2^\pm=\frac{1}{2}~.
\label{alpha-final}
\end{equation}
The general expressions for the corresponding coefficients $\gamma_{1,2}^{\pm}$ are also given in \autoref{power-law-relaxation}. For $r<2$, we evaluate $\gamma_{1,2}^{\pm}$ as an explicit example.

In fact, the exponents (\ref{alpha-final}) are quite generic. For $\Omega_{0\pm} = \Omega_{1,2}$ one has $\frac{\partial J}{\partial \theta} \sim \frac{\partial \Omega}{\partial \theta} = 0$. Assuming that the second derivative is non-vanishing, a Taylor expansion yields the form $\Omega(\theta_{0\pm}) \sim \Omega(\theta_{1,2}) + \kappa^{\pm}_{1,2} [(\theta_{0,\pm} - \theta (\Omega_{1,2})]^2 + \cdots$ which immediately leads to the result. A less generic situation involves vanishing second derivatives leading to exponents $1/4$ or smaller. See \cite{mandalkulkarni} for a discussion about this point.

\section{Conclusion}
\label{secsix}

Our results show that the space-like singularities found in geometries emerging from time dependent solutions of the $c=1$ matrix quantum mechanics are no longer present once we depart from the strict double scaling limit. In the fermion phase space these solutions correspond to fermions draining off at infinity, and a proper understanding of the late time state cannot be possible without an IR completion. In this paper we used a quartic potential as the IR completion and showed that at time scales of $[O(\log N)]^2$ the fermion system approaches a Generalized Gibbs Ensemble. The details of this approach of course depend on the particular IR completion and are not universal. However the fact that such a steady state is approached is clear.

It is important to note that the phase space density keeps oscillating in time around some background function at arbitrarily late times. However in terms of a coarse grained density obtained by averaging over an energy window the system relaxes to a steady state in a power law fashion with universal exponents.

The main question is: what is the String Theory interpretation of this state? Note that we are working in a regime where the unscaled fermi level $\mu_F$ is not tuned to its critical value. Consequently, if we think of the 't Hooft Feynman diagrams generated by the matrix model as discretized world sheets, we are away from the limit where the worldsheet is continuous. Intriguingly, it has been recently suggested that matrix integrals can have a string interpretation even away from the double scaling limit \cite{gopa, mazenc} -- it would be interesting if there is a similar correspondence in our case. We hope to return to this important issue in the near future.

\section*{Acknowledgements} S.R.D would like to thank A. Jevicki and T. Takayanagi for discussions at various stages of this work. The work of S.R.D. is partially supported by National Science Foundation grant NSF-PHY/2410647. The work of S.D.H is supported by KIAS Grant PG096301. S.R.D and G.M would like to thank the Banff International Research Station (BIRS) workshop ``Large N Matrix Models and Emergent Geometry" (26w5512) where part of this work was done.

\appendix

\section{Double Scaling Limit}
\label{doubscale}

The action of this model is given by
\ben
S = \beta \int dt ~{\rm Tr} \left[ (D_t M)^2 - U(M) \right]~, \qquad \beta = \frac{N}{\kappa}~.
\label{0-2a}
\een
Here $M(t)$ is a $N \times N$ Hermitian matrix, $\beta \sim N$ is a coupling and $D_t = \partial_t -iA_t$ is a covariant derivative. $U(M)$ is a potential which has a quadratic maximum at $M=0$, e.g.
\ben
U(M) = -\frac{1}{2}M^2 + \frac{1}{4} M^4 + \cdots
\label{0-3a}
\een
As is well known, in the $A_t=0$ gauge the Gauss law constraint restricts us to the singlet sector. The theory then reduces to the theory of eigenvalues $\lambda_i, i = 1 \cdots N$. After a standard redefinition of the wavefunction, these eigenvalues become free fermions. The Hamiltonian of the second quantized fermion field $\Psi (\lambda,t)$
\ben
H_F = \int d\lambda \left[ \frac{1}{2\beta} (\partial_\lambda \Psi^\dagger)
(\partial_\lambda \Psi) + \beta U(\lambda) \Psi^\dagger \Psi \right]~, \qquad \int d\lambda ~\Psi^\dagger \Psi = N~.
\label{0-3b}
\een
The double scaling limit is $N \rightarrow \infty$ and the fermi energy $-\epsilon_F \rightarrow 0$ with 
\ben
g_s =- \frac{1}{2 \beta \mu_F} 
\label{0-4a}
\een
and $\kappa$ held fixed. The limit $\mu_F \rightarrow 0$ renders the worldsheet formed by Feynman-'t Hooft diagrams continuous. Then $g_s$ appears as the string coupling providing weights of worldsheets with different topologies.
Rescaling coordinates and fields 
\ben
\lambda = (\beta g_s)^{-1/2} x~, \qquad \Psi = (\beta g_s)^{1/4} \psi~,
\label{0-5a}
\een
one immediately sees that only the universal quadratic part of the potential survives in the double scaling limit. The higher order terms in $U(\lambda)$ are suppressed by powers of $1/(\beta g_s)$. The double scaled fermion hamiltonian becomes
\ben
H_{ds} =\frac{1}{g_s} \int dx \left[ \frac{g_s^2}{2} (\partial_x \psi^\dagger)
(\partial_x \psi) + \frac{1}{2} U(x) \psi^\dagger \psi \right]~, \qquad \int dx ~\psi^\dagger \psi = N~.
\label{0-6a}
\een
The collective field and its conjugate momentum are given by the operator relations
\ben
\rho (\lambda) = \frac{1}{\beta} \Psi^\dagger \Psi =  \frac{1}{\beta} \delta (\lambda \cdot I - M)~, \qquad \Pi_\rho = -i \frac{\delta}{\delta \rho}~.
\label{0-7b}
\een
We now rescale
\ben
\rho = (\beta g_s)^{-1/2}\partial_x \varphi~, \qquad \partial_\lambda\Pi_\rho = -(\beta g_s)^{3/2} \Pi_\varphi
\label{0-8b}
\een
to pass to the double scaling limit. At leading order, the Hamiltonian for the collective field is given by
\ben
H = \int dx \left[ \frac{g_s^2}{2} \Pi_\varphi (\partial_x \varphi) \Pi_\varphi + \frac{\pi^2}{6g_s^2} (\partial_x \varphi)^3- \frac{x^2}{2g_s^2} (\partial_x\varphi)  + \frac{1}{2g_s^2} (\partial_x\varphi) \right]~.
\label{0-9b}
\een
The last term comes from a Lagrange multiplier which imposes the condition $\int dx \psi^\dagger \psi = N$. 

\section{Calculation of $E_{\max}$, $E_{\min}$, and the corresponding winding time}
\label{maxmin}

\subsection{Period of individual fermions in the double-well potential}

To obtain the period, it is useful to rewrite the Jacobi elliptic functions in terms of the Jacobi amplitude,
\begin{equation}
\operatorname{sn}(u,k)=\sin[\operatorname{am}(u,k)]~,
\qquad
\operatorname{cn}(u,k)=\cos[\operatorname{am}(u,k)]~,
\qquad
\operatorname{dn}(u,k)=
\sqrt{1-k^2\sin^2[\operatorname{am}(u,k)]}~.
\end{equation}
Since
\begin{equation}
\operatorname{am}(u+2K(k),k)=\operatorname{am}(u,k)+\pi~,
\end{equation}
the orbit returns to itself after $u\to u+2K$. Therefore,
\begin{equation}
T(-E)=2K(k)\omega^{-1}~.
\tag{\ref{1-15}}
\end{equation}
The same result can also be obtained from the standard period integral representation, as shown in \eqref{1-13}-\eqref{1-15}.

The logarithmic divergence of the period follows from the standard asymptotic expansion of $K(k)$ in the limit $E \rightarrow0+$. Expanding at small $ g E$ gives
\begin{equation}
T(-E)
=
\log(64)-\log(4g E)
+
\frac{1}{16}(4g E)
\left[
-3\log(4g E)-10+18\log2
\right]
+
\mathcal O\!\left(
(4g E)^2
\right)~.
\label{period-series}
\end{equation}
This energy dependence will be used below to estimate the period difference and, from it, the time scale for the deformed Fermi surface to wind once.

\subsection{$E_{\max}$ and $E_{\min}$ on the deformed Fermi surface}

Consider the deformed Fermi surface
\begin{equation}
x^2-p^2-\frac12 gx^4-|x-p|^r
=
2\mu~ .
\label{fs_deform}
\end{equation}
The extremal-energy trajectories,
\begin{equation}
x^2-p^2-\frac12 gx^4
=
2E~, \qquad E>0~,
\end{equation}
are tangent to the deformed Fermi surface. Therefore, at the tangent point, the two curves have the same energy and the gradients of their defining functions must be parallel. This gives
\begin{equation}
p_0=x_0-gx_0^3~.
\label{extE_rel}
\end{equation}

Substituting (\ref{extE_rel}) into (\ref{fs_deform}) gives
\begin{equation}
\frac32 g x_0^4
-
g^2 x_0^6
-
g^r |x_0|^{3r}
=
2\mu ~.
\label{constraint}
\end{equation}
It is convenient to introduce
\begin{equation}
4gE
=
4g\mu+2y ~.
\label{def_y}
\end{equation}
Using
\begin{equation}
2y
=
2g|x_0-p_0|^r
=
2g^{1+r}|x_0|^{3r}~,
\end{equation}
(\ref{constraint}) becomes
\begin{equation}
H(y)
=
\frac32 g^{-1/3}
\left(g^{-1}y\right)^{\frac{4}{3r}}
-
\left(g^{-1}y\right)^{\frac{2}{r}}
-
g^{-1}y
-
2\mu
=
0~.
\end{equation}

Although the roots cannot be obtained analytically, their small-$g$ scaling can be estimated from the leading terms of $H(y)$. This gives the smaller root
\begin{equation}
y_{\min}
\sim
\left(
\frac{4\mu}{3}
\right)^{\frac{3r}{4}}
g^{1+\frac r4}~,
\end{equation}
and the larger root
\begin{equation}
y_{\max}
\sim
\begin{cases}
\left(\frac32\right)^{\frac{3r}{2}}
g^{\frac{2-r}{2}}~,
&
r\le2~,
\\[6pt]
\left(\frac32\right)^{\frac{3r}{3r-4}}
g^{\frac{2r-4}{3r-4}}~,
&
r>2~.
\end{cases}
\end{equation}
According to the definition of $y$ in \eqref{def_y}, the extremal values are
\begin{equation}
4gE_{\min}
=
4g\mu+2y_{\min}~,
\qquad
4gE_{\max}
=
4g\mu+2y_{\max}~.
\end{equation}
For small $g$, $y_{\min}\ll g\mu$ while $y_{\max}\gg g\mu$ in the regime of interest. Hence
\begin{equation}
4gE_{\min}\sim 4g\mu~,
\qquad
4gE_{\max}\sim 2y_{\max}~.
\end{equation}

\subsection{Time scale for winding once}

We first estimate the period difference between the extremal trajectories,
\begin{equation}
\Delta T
=
 T(-E_{\min})
-
 T(-E_{\max}) ~.
\end{equation}
Substituting the extremal energies into \eqref{period-series} yields for $E \to 0+$ and small $g$
\begin{equation}
\Delta T
\sim
\log
\frac{4gE_{\max}}
     {4gE_{\min}}  \sim -\log(2\mu)
+
\frac{r}
     {\max\{3r-4,\,2\}}
\log
\left(
\frac{27}{8g}
\right)~.
\end{equation}

We introduce a typical time scale, the winding time, as the time needed for the fastest orbit, namely the orbit with energy $-E_{\max}$, to get one period ahead of the slowest orbit, with energy $-E_{\min}$. This gives
\begin{equation}
T_{\rm winding}
=
\frac{2\pi}
{
\frac{2\pi}{ T(-E_{\max})}
-
\frac{2\pi}{ T(-E_{\min})}
}
=
\frac{
T(-E_{\min})T(-E_{\max})
}
{
\Delta  T
}~.
\end{equation}
Using the leading logarithmic behaviour of (\ref{period-series}), one finds
\begin{equation}
T_{\rm winding}
\sim
\left[
\log(4g\mu)
\right]^2~.
\end{equation}

Hence the detailed energy range depends on the deformation exponent $r$, whereas the leading winding time is universal and grows as $(\log g )^2$ in the weak-coupling limit.

\section{Action-angle variables in the double-well potential}
\label{action-angle}

In this appendix we summarize the action-angle variables for the one-particle motion in the double-well potential. 

Consider the trajectory of individual fermions with energy $-E$ given in \eqref{1-11}. Without loss of generality, we choose the $x>0$ branch, in which a fermion oscillates between two positive turning points $c$ and $a$, defined by
\begin{equation}
a^2 = \frac{1}{g}(1+\sqrt{1-4gE})~,
\qquad
c^2 = \frac{1}{g}(1-\sqrt{1-4gE})~.
\label{a_and_c}
\end{equation}
The modulus $k$ in \eqref{1-10} is then simply determined by
\begin{equation}
k^2 = 1- \frac{c^2}{a^2}~.
\end{equation}

The action variable is
\begin{equation}
J
=
\frac{1}{2\pi}
\oint p~dx~ .
\end{equation}
With the counterclockwise orientation of the closed orbit, the contribution on the upper branch runs from $x=a$ to $x=c$. Therefore,
\begin{equation}
J
= \frac{1}{\pi}
\int_a^c \sqrt{-2E + x^2 - \frac{g}{2} x^4}~ dx = 
\frac{1}{\pi}
\int_a^c
\sqrt{\frac{g}{2}(a^2-x^2)(x^2-c^2)}~dx~ .
\end{equation}
The integral can be evaluated in terms of complete elliptic integrals, yielding
\begin{equation}
J
=
-\frac{\sqrt{g/2}~a}{3\pi}
\left[
-2c^2 \bK(k)
+
(a^2+c^2) \bE(k)
\right],
\end{equation}
where $\bK(k)$ and $\bE(k)$ are the complete elliptic integrals of the first and second kinds. In terms of the frequency parameter $\omega$ defined in \eqref{1-10}, we recover \eqref{5-1}.

Since the transformed Hamiltonian depends only on $J$, the angle $\theta$ evolves linearly,
\begin{equation}
\dot{\theta}
=
\frac{dH}{dJ}
=
\left(
\frac{dJ}{d (-E)}
\right)^{-1}~.
\end{equation}
Differentiating the action with respect to the energy gives
\begin{equation}
\frac{dJ}{d(-E)}
=
\frac{1}{\pi}
\int_a^c
\frac{dx}
{\sqrt{-2E + x^2 - \frac{g}{2} x^4}} ~.
\end{equation}
Utilising the conclusion in \eqref{1-15}, we immediately find
\begin{equation}
\frac{dJ}{d(-E)}
=
-\frac{\bK(k)}{\pi\omega},
\qquad
\omega
=
\sqrt{\frac{g}{2}}~a ~.
\end{equation}
Therefore
\begin{equation}
\dot{\theta}
=
-\frac{\pi\omega}{\bK(k)} = \Omega (J) ~ .
\end{equation}
Integrating both sides gives \eqref{5-5}.

The same angle variable can be obtained from the generating function
\begin{equation}
W(x,J)
=
\int^x p(x',J)\,dx',
\qquad
\theta
=
\frac{\partial W}{\partial J}~.
\end{equation}
Note that $x$ is a function of time $t$, so the integral should be treated as a contour integral along the classical trajectory, with the orientation determined by the time evolution.
For the right-well branch $x>0$, and choosing the origin of the angle at the outer turning point,
\begin{equation}
\theta(a,0)=0~,
\end{equation}
one finds
\begin{equation}
\theta
=
\begin{cases}
\displaystyle
\frac{\pi}{\bK(k)}
\bF\!\left( \arcsin
\sqrt{\frac{a^2-x^2}{a^2-c^2}};k
\right)~,
&
p > 0~,
\\[12pt]
\displaystyle
\frac{\pi}{\bK(k)}
\left[
2K(k)
-
\bF\!\left( \arcsin
\sqrt{\frac{a^2-x^2}{a^2-c^2}};k
\right)
\right]~,
&
p < 0~,
\end{cases}
\end{equation}
where $\bF( \phi ;k)$ is the incomplete elliptic integral of the first kind. The two cases correspond to the two branches of the same closed orbit. The formula is continuous at the turning points, up to the usual identification $\theta\sim\theta+2\pi$.

For the left-well branch $x<0$, the result is obtained by the reflection
\begin{equation}
x\rightarrow -x~,
\qquad
p\rightarrow -p~ .
\end{equation}
Thus the two wells are described by the same action variable, while the angle variable keeps track of the position along each classical orbit.

Combining the angle variables of the two branches yields the angle variable in \eqref{5-1}.

\section{Late-Time Power-Law Relaxation} 
\label{power-law-relaxation}

In this appendix, we determine the exponents and coefficients of the power-law relaxation $\alpha_{1,2}^{\pm}$ and $\gamma_{1,2}^{\pm}$ appearing in \eqref{5-15} and \eqref{5-17}.
Throughout this appendix, we work in the regime
\begin{equation}
    4gE \ll 1.
\end{equation}

\subsection{Intersections of Static Fermion Orbits with Near Extremal Energies
and Initial Fermi Surface}

The angle variables $\theta_{0,\pm}$ are evaluated at the two intersections
of the constant energy orbit in \eqref{1-11}, which is also a constant-$J$ orbit, with the initial Fermi surface
\begin{equation}
     -p^2 +x^2 -\frac{g}{2}x^4 - |x-p|^r =1~.
\label{app:initial-fermi-surface}
\end{equation}
Note that \eqref{app:initial-fermi-surface} corresponds to $F(x,p)=1$ in \eqref{1-8}, with the initial time chosen to be $t=-T=0$.
The extremal intersections, denoted as $(x_0, p_0)$, have already been considered in \autoref{maxmin} and the corresponding maximal or minimal energies $E_{\max}$ and $E_{\min}$ have been calculated for small $g$. According to the explanation below \eqref{5-12}, these extremal energies determine the corresponding angle variables
\begin{equation}
\theta (\Omega_1) = \theta (\Omega(J(E_{\min})))~, \qquad  \theta (\Omega_2) = \theta (\Omega(J(E_{\max})))~,
\label{app:redef-Omega}
\end{equation}
around which we expand $\theta_{0,\pm}$ below.

Now we consider a nearby static orbit with energy $E= E_0+\Delta E$, where $E_0$ denotes either $E_{\max}$ or $E_{\min}$, and write
\begin{equation}
    x=x_0+\Delta x, \qquad p=p_0+\Delta p.
\end{equation}
Expanding \eqref{1-11} around $(x_0,p_0)$ gives
\begin{equation}
\begin{split}
    -2\Delta E = & 2(x_0-gx_0^3)(\Delta p-\Delta x) +(\Delta p-\Delta x)^2 \\
    &+ 2(\Delta p-\Delta x)\Delta x +3gx_0^2(\Delta x)^2 +\mathcal{O}\left((\Delta x)^3\right)~,
\end{split}
    \label{app:static-expansion}
\end{equation}
where we have utilised the constraint on the extremal intersections \eqref{1-12a}.

By symmetry, it is sufficient to consider the positive-$x$ branch. For a nearby intersection,
\begin{equation}
    |x-p| =\left| gx_0^3+\Delta x-\Delta p \right|~.
\end{equation}
Now if we assume
\begin{equation}
    |\Delta x-\Delta p|\ll gx_0^3,
\end{equation}
the quantity inside the absolute value remains positive, and hence the initial Fermi-surface condition gives
\begin{equation}
    2\Delta E = \left( gx_0^3+\Delta x-\Delta p  \right)^r -(gx_0^3)^r.
\end{equation}
For
\begin{equation}
    \frac{|\Delta E|}{(gx_0^3)^r}\ll1,
    \label{app:intersection-validity}
\end{equation}
we obtain
\begin{equation}
    \Delta x-\Delta p = \frac{2\Delta E}{r} (gx_0^3)^{1-r} +\mathcal{O}\left((\Delta E)^2\right).
    \label{app:dx-dp-general}
\end{equation}
Indeed, \eqref{app:dx-dp-general} implies
\begin{equation}
    \frac{|\Delta x-\Delta p|}{gx_0^3} \simeq \frac{2|\Delta E|}{r(gx_0^3)^r} \ll1,
\end{equation}
hence the removal of the absolute value above is self-consistent.

Equation \eqref{app:dx-dp-general} fixes $\Delta p-\Delta x$ in terms of $\Delta E$. Plugging it into \eqref{app:static-expansion}, we obtain an equation involving only $\Delta x$,
\begin{equation}
0=3gx_0^2(\Delta x)^2 -\frac{4\Delta E}{r}(gx_0^3)^{1-r}\Delta x 
+2\Delta E \left[ 1-\frac{2}{r}(x_0-gx_0^3)(gx_0^3)^{1-r} \right]+\cdots .
\label{app:quadratic-dx}
\end{equation}
A consistent solution is given by $(\Delta x)^2\sim |\Delta E|$, so that the term proportional to $\Delta E\,\Delta x$ is subleading. Therefore,
\begin{equation}
(\Delta x)^2 \simeq \frac{2\Delta E}{3gx_0^2} \left[ \frac{2}{r}(x_0-gx_0^3)(gx_0^3)^{1-r}-1 \right] = C_x (E_0,g,r)^2 |\Delta E|~.
\label{app:dx-square-general}
\end{equation}
For the values of $E$ for which the nearby intersections exist, the right-hand side of \eqref{app:dx-square-general} is positive, and the two intersections therefore behave as
\begin{equation}
\Delta x_{\pm}= \pm C_x(E_0,g,r) |\Delta E|^{1/2}+\mathcal{O}(|\Delta E|)~.
\label{app:dx-sqrt-general}
\end{equation}
Consequently,
\begin{equation}
gx_{\pm}^2-gx_0^2 = \pm C_{x^2}(E_0,g,r) |\Delta E|^{1/2} +\mathcal{O}(|\Delta E|)~,
\label{app:gx-square-general}
\end{equation}
where
\begin{equation}
C_{x^2}(E_0,g,r) = 2 gx_0 \cdot  C_x(E_0,g,r) =  2\left(  \frac{2}{3 } \right)^{1/2}   \left|   \frac{2}{r} \frac{(g x_0^2)^2 \left( 1 -gx_{0}^2 \right)}{ (gx_{0}^3)^{r}}  -g \right|^{1/2}~.
\end{equation}

The condition \eqref{app:intersection-validity} ensures that the higher-order corrections in $\Delta E$ remain smaller than the leading terms in the near-extremal-energy expansion.

\subsection{Asymptotic Behaviour of the Angle Variable}

We next show that the square-root splitting \eqref{app:gx-square-general} directly determines the relaxation exponents $\alpha_{1,2}^{\pm}$.

According to \eqref{5-1} (and also \eqref{5-2}), in order to determine the asymptotic behaviour of $\theta_{0,\pm}$ around the extremal energies, we need to expand $\bF(\phi,k)$ and $\bK(k)$ near the corresponding extremal intersections. We also note that the sign of $p_0\operatorname{sgn}(x_0)$ determines which branch of $\theta(x,p)$ should be used for each extremal intersection.

We first examine the explicit energy dependence of $k$ and $\omega$. From \eqref{1-10}, we find that for $E=E_0+\Delta E$, expanding $k(E)^2$ and $\omega(E)^2$ around $E_0$ gives
\begin{equation}
\omega(E)^2-\omega(E_0)^2= -\frac{g}{\sqrt{1-4 g E_0}} \Delta E +\mathcal{O}\left((\Delta E)^2\right)~,
\end{equation}
and
\begin{equation}
k(E)^2-k(E_0)^2 = -\frac{4g}{\sqrt{1-4g E_0}\left(1+\sqrt{1-4gE_0}\right)^2}\Delta E +\mathcal{O}\left((\Delta E)^2\right)~.
\label{app:k-expansion}
\end{equation}
Thus, the explicit energy dependence of $k$ and $\omega$ produces corrections of order $\Delta E$.

We now consider the corresponding variation of $\lambda = \sin \phi$. According to \eqref{5-2},
\begin{equation}
\begin{split}
\lambda^2(E,x_\pm) =&\frac{1-\frac{1}{2}\left[gx_0^2+ \left(gx_\pm^2-gx_0^2\right)\right]\omega^{-2}(E)}{k^2(E)}\\
=& \frac{1-\frac{1}{2}gx_0^2\omega^{-2}(E)}{k^2(E)}+\frac{1}{2}\left(gx_0^2-gx_\pm^2\right) \omega^{-2}(E)k^{-2}(E)~.
\end{split}
\label{app:lambda-direct-expansion}
\end{equation}
The first term differs from $\lambda^2(E_0,x_0)$ only through the explicit energy dependence of $k$ and $\omega$, and therefore
\begin{equation}
\frac{1-\frac{1}{2}gx_0^2\omega^{-2}(E)}{k^2(E)}= \frac{1-\frac{1}{2}gx_0^2 \left( \omega^{-2}(E_0) +\mathcal{O} (g \Delta E) \right)}{k^2(E_0)+\mathcal{O} (g \Delta E) } = \lambda^2(E_0,x_0)+\mathcal{O}(\Delta E)~.
\end{equation}
Furthermore, \eqref{1-10} shows
\begin{equation}
\omega^2(E)k^2(E)=\sqrt{1-4g E}~,
\end{equation}
so that, for $4gE \ll1$,
\begin{equation}
\omega^{-2}(E)k^{-2}(E)=\frac{1}{\sqrt{1-4gE}}=1+\frac{4gE}{2}+\mathcal{O}((gE)^2)
\label{app:omega-modulus-prod-expansion}
\end{equation}
only gives a relative correction of order $gE$ to the leading $|\Delta E|^{1/2}$ term.
\eqref{app:gx-square-general}, however, gives
\begin{equation}
gx_\pm^2-gx_0^2=\mathcal{O}\left(|\Delta E|^{1/2}\right)~.
\end{equation}
Therefore,
\begin{equation}
\lambda^2(E,x_\pm)= \lambda^2(E_0,x_0) +\frac{1}{2}\left(gx_0^2-gx_\pm^2\right)+\mathcal{O}(|\Delta E|)+\mathcal{O}\left( gE|\Delta E|^{1/2}\right)~.
\label{app:lambda-square-general-expansion}
\end{equation}
Taking the square root of \eqref{app:lambda-square-general-expansion}, we obtain
\begin{equation}
\lambda(E,x_\pm)- \lambda(E_0,x_0) = \frac{ gx_0^2-gx_\pm^2 }{ 4\lambda(E_0,x_0)  } +\mathcal{O}(|\Delta E|) +\mathcal{O}\left( gE |\Delta E|^{1/2} \right)~.
\label{app:lambda-general-expansion}
\end{equation}
In conclusion, we have
\begin{equation}
\lambda(E,x_\pm)- \lambda(E_0,x_0)=\mathcal{O}\left(|\Delta E|^{1/2}\right)~.
\label{app:lambda-sqrt}
\end{equation}

We should also make sure that, as the energy is varied away from $E_0$, the nearby intersections do not cross between the two branches of $\theta(x,p)$. Since the allowed range of $\lambda$ is $0\leq\lambda\leq1$, with
\begin{equation}
\begin{cases}
\lambda=0 &\Longleftrightarrow gx^2=1+\sqrt{1- 4gE}~, \\
\lambda=1 &\Longleftrightarrow gx^2=1-\sqrt{1-4gE}~,
\end{cases}
\end{equation}
according to \eqref{a_and_c}, the nearby intersections remain away from the branch boundaries provided
\begin{equation}
0<\lambda^2(E,x_\pm) \approx \lambda^2(E_0,x_0) +\frac{1}{2}\left(gx_0^2-gx_\pm^2\right)<1~.
\end{equation}
Using \eqref{5-2} and \eqref{app:gx-square-general}, this condition gives a constraint on the allowed size of $\Delta E$ relative to the extremal value $E_0$ and $gx_0^2$
\begin{equation}
-\frac{1+\sqrt{1-4gE_0}}{\sqrt{1-4gE_0}} +\frac{gx_0^2}{\sqrt{1-4gE_0}}
< C_{x^2}|\Delta E|^{1/2}
< \frac{-1+\sqrt{1-4gE_0}}{\sqrt{1-4gE_0}}+\frac{gx_0^2}{\sqrt{1-4gE_0}}~.
\label{app:branch-condition-general}
\end{equation}
For $gE_0\ll1$, a sufficient condition is therefore
\begin{equation}
2\left(  \frac{2}{3 } \right)^{1/2}   \left|   \frac{2}{r} \frac{(g x_0^2)^2 \left( 1 -gx_{0}^2 \right)}{ (gx_{0}^3)^{r}}  -g \right|^{1/2} |\Delta E|^{1/2} < \min\left\{gx_0^2, 2-gx_0^2 \right\}~,
\label{app:branch-condition-sufficient}
\end{equation}
which is equivalent to
\begin{equation}
\frac{|\Delta E|}{(gx_{0}^3)^{r}} < 
\frac{3}{8}
\begin{cases}
\frac{g^2x_0^4}{\left|   \frac{2}{r} \left[ (g x_0^2)^2 \left( 1 -gx_{0}^2 \right) \right]  -g (gx_{0}^3)^{r} \right| } ~, & gx_0^2 <1~,\\
\frac{(2-gx_0^2)^2}{\left|   \frac{2}{r} \left[ (g x_0^2)^2 \left( 1 -gx_{0}^2 \right) \right]  -g (gx_{0}^3)^{r} \right| } ~, & gx_0^2 > 1~
\label{app:no-crossing-cond}
\end{cases}
~.
\end{equation}
Using \eqref{constraint}, we can find 
\begin{equation}
g (gx_{0}^3)^{r}  < (gx_0^2)^2 \left(  \frac{3}{2} - gx_0^2\right)
\end{equation}
and hence $0 < gx_0^2 < \frac{3}{2}$ from the positivity of $gx_0^3$. As a result, for fixed $r>0$, the right-hand side of \eqref{app:no-crossing-cond} can be bounded from below by a positive $r$-dependent number of order 1.\footnote{For $gx_0^2 <1$, we can bound the denominator by 
\begin{equation}
\left|   \frac{2}{r} \left[ (g x_0^2)^2 \left( 1 -gx_{0}^2 \right) \right]  -g (gx_{0}^3)^{r} \right| < \frac{2}{r} (g x_0^2)^2 + \frac{3}{2} (g x_0^2)^2~,
\end{equation}
while for $gx_0^2 >1$, the numerator is bounded by $(2-gx_0^2)^2 > \frac{1}{4}$ and the denominator is bounded by 
\begin{equation}
\left|   \frac{2}{r} \left[ (g x_0^2)^2 \left( 1 -gx_{0}^2 \right) \right]  -g (gx_{0}^3)^{r} \right| < \frac{2}{r} \left(\frac{3}{2}\right)^2 \left( \frac{3}{2} -1 \right) + 1^2 \left(  \frac{3}{2} - 1\right) = \frac{9}{4r} + \frac{1}{2}
\end{equation}
where the first and second terms are maximised at $gx_0^2 =\frac{3}{2}$ and $gx_0^2 = 1$, respectively.} Therefore, the near-extremal condition \eqref{app:intersection-validity} is sufficient to ensure that the nearby intersections do not cross the branch boundaries of $\theta(x,p)$.

The incomplete elliptic integral of the first kind can be written as
\begin{equation}
\bF(\lambda;k)=\int_0^\lambda\frac{dt}{\sqrt{1-t^2}\sqrt{1-k^2t^2}}~.
\end{equation}
Near an extremal intersection, both $\lambda(E,x_\pm)$ and $k(E)$ vary with the energy. Expanding first in $\lambda$ and then in $k$ around the extremal point, we obtain
\begin{equation}
\begin{split}
\bF(\lambda(E,x_\pm);k(E))
=& \bF(\lambda(E_0,x_0);k(E_0)) + \frac{\partial\bF(\lambda(E_0,x_0);k)}{\partial k} \bigg|_{k=k(E_0)} \left[k(E)-k(E_0)\right] \\
&+\frac{\lambda(E,x_\pm)-\lambda(E_0,x_0)}{\sqrt{1-\lambda(E_0,x_0)^2}\sqrt{1-k(E)^2\lambda(E_0,x_0)^2}}
+\cdots~.
\end{split}
\label{app:F-general-expansion}
\end{equation}
The derivative is evaluated at the extremal point and is therefore independent of $\Delta E$, while $k(E)-k(E_0)=\mathcal{O}(\Delta E)$. Therefore, the variation of $\bF$ through its explicit $k$ dependence contributes only at order $\Delta E$. A further expansion of the coefficient of $\lambda(E,x_\pm)-\lambda(E_0,x_0)$ around $k(E_0)^2$ gives a correction proportional to $k(E)^2-k(E_0)^2$, which is of order $\Delta E$ according to \eqref{app:k-expansion}. Thus, the leading $|\Delta E|^{1/2}$ contribution is determined entirely by the variation of $\lambda$, since \eqref{app:lambda-general-expansion} and \eqref{app:gx-square-general} give the leading square-root contribution
\begin{equation}
\lambda(E,x_\pm)-\lambda(E_0,x_0) = \mp \frac{ C_{x^2}(E_0,g,r) }{ 4\lambda(E_0,x_0) } |\Delta E|^{1/2} +\mathcal{O}(|\Delta E|)~.
\end{equation}
We thus obtain
\begin{equation}
\begin{split}
\bF(\lambda(E,x_\pm);k(E)) 
=&\bF(\lambda(E_0,x_0);k(E_0)) \\
&\mp \frac{C_{x^2}(E_0,g,r)}{4\lambda(E_0,x_0)\sqrt{1-\lambda(E_0,x_0)^2} \sqrt{1-k(E_0)^2\lambda(E_0,x_0)^2}}|\Delta E|^{1/2} \\
&+\mathcal{O}(|\Delta E|)~.
\end{split}
\label{app:F-sqrt-expansion}
\end{equation}

Similarly, expanding $\bK(k(E))$ around $k(E_0)$, the derivative $d\bK/dk|_{k=k(E_0)}$ is independent of $\Delta E$, while $k(E)-k(E_0)=\mathcal{O}(\Delta E)$. Therefore, the variation of $\bK(k(E))$ starts at order $\Delta E$ and does not contribute to the leading $|\Delta E|^{1/2}$ term in the angle variable. Since the no-crossing condition established above (\eqref{app:branch-condition-general}-\eqref{app:no-crossing-cond}) ensures that the nearby intersections remain on the same branch of $\theta(x,p)$, \eqref{5-1} gives
\begin{equation}
\begin{split}
& \theta_{0,\pm}(E)-\theta(E_0) \\
& \qquad \simeq 
\begin{cases}
\displaystyle
\begin{split}
\mp\frac{\pi C_{x^2}(E_0,g,r)}{4\bK(k(E_0))\lambda(E_0,x_0)\sqrt{1-\lambda(E_0,x_0)^2}\sqrt{1-k(E_0)^2\lambda(E_0,x_0)^2}}|\Delta E|^{1/2}&~, \\[10pt]
p_0\operatorname{sgn}(x_0)& >0~,
\end{split}
\\[14pt]
\displaystyle
\begin{split}
\pm\frac{\pi C_{x^2}(E_0,g,r)}{4\bK(k(E_0))\lambda(E_0,x_0)\sqrt{1-\lambda(E_0,x_0)^2}\sqrt{1-k(E_0)^2\lambda(E_0,x_0)^2}}|\Delta E|^{1/2}& ~,\\[10pt]
p_0\operatorname{sgn}(x_0)&<0~.
\end{split}
\end{cases}
\end{split}
\label{app:theta-energy-general}
\end{equation}

We next express the energy deviation in terms of the angular-velocity deviation. Expanding $|\Omega(E)|$ around $E_0$, we have
\begin{equation}
|\Omega(E)|-|\Omega(E_0)|
=  \frac{\partial|\Omega|}{\partial E}\bigg|_{E_0} (E-E_0)+\mathcal{O}\left((E-E_0)^2\right).
\label{app:Omega-linear-general}
\end{equation}
Therefore,
\begin{equation}
|\Delta E|^{1/2} = \left[\frac{\partial|\Omega|}{\partial E}\bigg|_{E_0}\right]^{-1/2}
\left||\Omega|-\Omega_0 \right|^{1/2} +\cdots,
\label{app:Omega-linear-general-inverse}
\end{equation}
where $\Omega_0 = |\Omega(E_0)|$ denotes $\Omega_1$ or $\Omega_2$, depending on the choice of $E_0$, as defined in \eqref{app:redef-Omega}. 

Moreover, \eqref{5-7} gives
\begin{equation}
\frac{\pi}{\bK(k(E_0))}=\frac{|\Omega(E_0)|}{\omega(E_0)}~.
\label{app:K-Omega-relation}
\end{equation}
Combining \eqref{app:Omega-linear-general-inverse} and \eqref{app:K-Omega-relation}
and plugging them into \eqref{app:theta-energy-general}, we find that the leading square-root contribution to the angle variable is proportional to
\begin{equation}
\frac{C_{x^2}(E_0,g,r)}{4\omega(E_0)\lambda(E_0,x_0)\sqrt{1-\lambda(E_0,x_0)^2}\sqrt{1-k(E_0)^2\lambda(E_0,x_0)^2}}\left[-\frac{\partial |\Omega|^{-1}}{\partial E}\bigg|_{E=E_0}\right]^{-1/2}\left|\Omega-\Omega_0\right|^{1/2}~,
\label{app:gamma-general-coefficient}
\end{equation}
where we have denoted $|\Omega|$ simply by $\Omega$.

Thus, comparing with \eqref{5-15} and \eqref{5-17}, we can read off
\begin{equation}
\alpha_1^\pm=\alpha_2^\pm=\frac{1}{2}
\tag{\ref{alpha-final}}
\end{equation}
from \eqref{app:gamma-general-coefficient}.
The coefficients $\gamma_{1,2}^{\pm}$ are then obtained by evaluating
\eqref{app:gamma-general-coefficient}, including the corresponding branch
sign, at $E_{\min}$ and $E_{\max}$. 

For completeness, the complete elliptic integral has the asymptotic expansion as $k\to 1$, 
\begin{equation}
\bK(k)=\sum_{m=0}^{\infty} \frac{\left(\tfrac{1}{2}\right)_m^2}{(m!)^2}{(1-k^2)}^{m}
\left[\ln\left(\frac{1}{\sqrt{1-k^2}}\right)+d(m)\right],
\qquad
d(m)=\psi(1+m)-\psi\left(\tfrac{1}{2}+m\right)~,
\end{equation}
where $\psi(z)$ denotes the digamma function.
Keeping the leading and subleading terms in this expansion, and expanding
$\omega(E)$ and $k(E)$ in \eqref{1-10} with respect to $4gE$, and neglecting
terms further suppressed by inverse logarithms, we obtain
\begin{equation}
\Omega(E) \simeq-\frac{2\pi}{\log\left(\frac{64}{4gE}\right)}\left(1-\frac{3gE}{4}\right).
\label{app:Omega-asymp}
\end{equation}
It follows that
\begin{equation}
-\frac{\partial |\Omega|^{-1}}{\partial E}\simeq
\frac{1}{2\pi E}\left[1-\frac{3gE}{4}
\log\left(\frac{64}{4gE}\right)
+\mathcal{O}(gE)\right].
\label{app:Omega-inverse-derivative}
\end{equation}

\subsubsection{Coefficients $\gamma_{1,2}^{\pm}$ for $r<2$}

As an explicit example, we now evaluate the coefficients $\gamma_{1,2}^{\pm}$ for $r<2$ using the general result \eqref{app:gamma-general-coefficient}. In this regime, the minimal and maximal energies were obtained in \autoref{maxmin}. Here we complete the extremal data by giving the corresponding values of $gx_0^2$:
\begin{equation}
(r<2) \quad 
\begin{cases}
4g E_{\min} =& 2g \left( 1+ \left(\frac{2}{3}\right)^{\frac{3r}{4}}g^{\frac r4} \right) \\
g x_{0,\min}^2 =& \sqrt{\frac{2}{3} g}
\end{cases} ~, \quad
\begin{cases}
4g E_{\max} =& 2 \left( \frac{3}{2} \right)^{\frac{3r}{2}} g^{\frac{2-r}{2}} \left( 1+ \left(\frac{2}{3}\right)^{\frac{3r}{2}}g^{\frac r2} \right) \\
g x_{0,\max}^2 =& \frac{3}{2}
\end{cases} ~.
\label{app:extremal-rless2}
\end{equation}
Using \eqref{1-12a}, these two endpoints satisfy
\begin{equation}
p_{0,\min}\operatorname{sgn}(x_{0,\min})>0,
\qquad
p_{0,\max}\operatorname{sgn}(x_{0,\max})<0,
\label{app:sign-rless2}
\end{equation}
and therefore correspond to the first and second branches of \eqref{app:theta-energy-general}, respectively.

Substituting the extremal data into \eqref{app:gx-square-general}, we find
\begin{equation}
\begin{split}
C_{x^2}(E_{\min},g,r)&\simeq 2 \sqrt{2}\left(\frac{2}{3}\right)^{1-3r/8}g^{1/2-r/8}\frac{1}{\sqrt{r}},
\\
C_{x^2}(E_{\max},g,r)&\simeq 2 \left(\frac{2}{3}\right)^{(3r-2)/4} g^{r/4} \frac{1}{\sqrt{r}}.
\end{split}
\label{app:Cx2-rless2}
\end{equation}
The corresponding values of $\lambda$ are
\begin{equation}
\lambda_{\min} \simeq 1-\frac{1}{4}\sqrt{\frac{2g}{3}},
\qquad
\lambda_{\max} \simeq \frac{1}{2} \left[ 1-\frac{1}{2} \left(\frac{3}{2}\right)^{3r/2} g^{(2-r)/2} \right].
\label{app:lambda-rless2}
\end{equation}
At $E_{\min}$,
\begin{equation}
1-\lambda_{\min}^2 \simeq \frac{1}{2}\sqrt{\frac{2g}{3}},
\qquad
1-k(E_{\min})^2=\mathcal{O}(g),
\end{equation}
so that
\begin{equation}
\sqrt{1-\lambda_{\min}^2} \sqrt{1-k(E_{\min})^2\lambda_{\min}^2}
\simeq
\frac{1}{2}\sqrt{\frac{2g}{3}}.
\end{equation}
At $E_{\max}$, on the other hand, $\lambda_{\max}\to 1/2$, $k(E_{\max})\to1$, and $\omega(E_{\max})\to1$, so all these factors approach finite constants.

Using \eqref{app:Omega-inverse-derivative}, we further obtain
\begin{equation}
\left[ -\frac{\partial |\Omega|^{-1}}{\partial E} \bigg|_{E=E_{\min}} \right]^{-1/2}
\simeq \sqrt{\pi} \left[1+\mathcal{O}\left(g^{r/4}\right) \right],
\end{equation}
and
\begin{equation}
\left[ -\frac{\partial |\Omega|^{-1}}{\partial E} \bigg|_{E=E_{\max}} \right]^{-1/2}
\simeq
\sqrt{\pi} \left(\frac{3}{2}\right)^{3r/4}g^{-r/4}\left[1+\mathcal{O}\left(g^{r/2}\right)+\mathcal{O}\left(g^{(2-r)/2}\log\frac{1}{g}\right)\right].
\end{equation}

Substituting these results into \eqref{app:gamma-general-coefficient}, together with the branch signs in \eqref{app:sign-rless2}, gives
\begin{equation}
\gamma_1^{\pm} = \pm \sqrt{2}\left(\frac{2}{3}\right)^{1/2-3r/8}
g^{-r/8}\left(\frac{\pi}{r}\right)^{1/2}\left[1+\mathcal{O}\left(g^{r/4}\right)\right],
\label{app:gamma1-final}
\end{equation}
and
\begin{equation}
\gamma_2^{\pm} = \mp 2 \left(\frac{2}{3}\right)^{1/2} \left(\frac{\pi}{r}\right)^{1/2}
\left[ 1+ \mathcal{O}\left(g^{r/2}\right) + \mathcal{O}\left(g^{(2-r)/2}\log\frac{1}{g} \right)\right].
\label{app:gamma2-final}
\end{equation}

Only the leading terms in the small-$g$ expansion are quantitatively controlled. The higher-order terms are retained only to indicate the parametric size of the corrections.

\section{Folds on the fermi surface}
\label{secfour}

Since a constant $x$ line intersects the fermi surface multiple times, there are many independent functions $P_{\pm , i}$ whereas at a classical level we have only two independent functions $\partial_x \varphi$ and $\Pi_\varphi$. If we insist of working with classical quantities we need more variables. A useful parametrization of a general fermi surface is in terms of $w_\infty$ charges \cite{avanjevicki}
\bea
\int dp~u(x,p) &=& \frac{1}{2\pi g_s} \left[ \beta_+ (x) - \beta_- (x) \right] \nonumber \\
\int dp~p~u(x,p) &=& \frac{1}{2\pi g_s} \left[\frac{1}{2}[ \beta_+ (x)^2 - \beta_- (x)^2] + [w_{1+} (x) -w_{1-}(x)] \right] \nonumber \\
\int dp~p^2~u(x,p) &=& \frac{1}{2\pi g_s} \left[\frac{1}{3}[ \beta_+ (x)^3 - \beta_- (x)^3] +[\beta_+ w_{1+} (x) -\beta_-w_{1-}(x)]  + [w_{2+} (x) - w_{2-}(x)] \right] ~.\nonumber \\
\label{4-1}
\eea
The fields $\beta_\pm, w_{i\pm}$ are, at the classical level canonically conjugate pairs, and their equations of motion follow from the Liouville equation for the phase space density. For a quadratic fermi surface $w_{i\pm} = 0$. If there are folds, or if a macroscopic number of fermions are excited leaving a gap, as in \autoref{folds-NR}. One can then choose
\ben
\beta_+=P_{1+}-P_{1-}+P_{2+}~, \qquad \beta_- = P_{2-}, \qquad w_{1+}= (P_{1+}-P_{1-})(P_{1-}-P_{2+})
~, \qquad w_{1-}=0 \cdots
\label{4-2}
\een

\begin{figure}[h]
\centering
\includegraphics[width=2.5in]{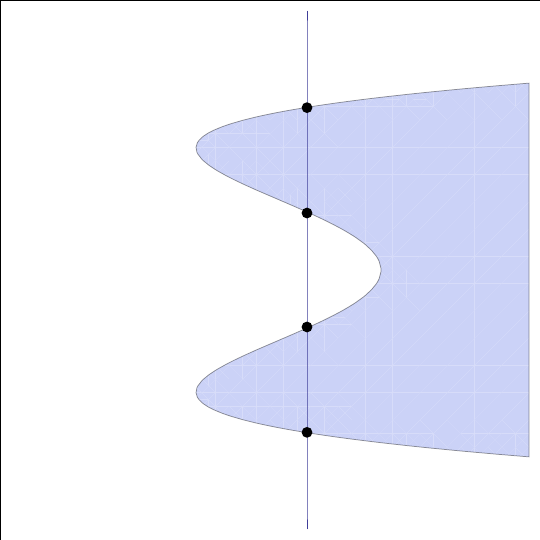}
\caption{A fold on the fermi surface for nonrelativistic fermions. The $x =$ constant line shown intersects the surface four times, labelled from the top by $P_{1+},P_{1-},P_{2+},P_{2-}$.}
\centering
\label{folds-NR}
\end{figure}
Nonzero $w_{i\pm}$ does not mean that a {\em quantum} bosonization in terms of a single field operator and its conjugate momenta is impossible \cite{fold1}. Rather, this means that quantum dispersions of the bosonic field are large, surviving the classical limit. 

This is clear when we have a relativistic chiral fermion where the operator relations are well known, and summarized in \autoref{appboson}. In this case the dispersion relation is linear, so that if we start with an initial state with no folds, a fold cannot develop at a later time. However, an initial state $|\chi\rangle$  itself can have a fold. The fermi sea is replaced by a Dirac sea, which has no bottom. This means that the formulae (\ref{4-1}) needs to be modified to reflect this fact. We can do this by simply omitting the $\beta_-, w_{i-}$. A typical fold in this case is shown in \autoref{folds-Chiral}) where a constant $x$ line intersects the surface thrice at $P_{1,\pm},P_{2+}$.
\begin{figure}[h]
\centering
\includegraphics[width=2.5in]{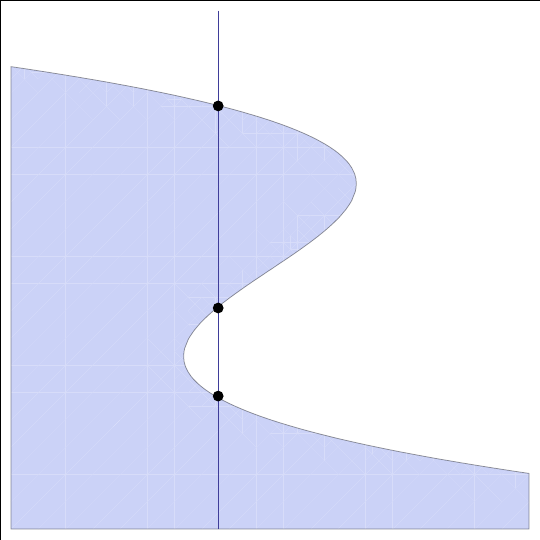}
\caption{A fold on the fermi surface for relativistic chiral fermion. The $x =$ constant line shown intersects the surface four times, labelled from the top by $P_{1+},P_{1-},P_{2+}$.}
\centering
\label{folds-Chiral}
\end{figure}
The standard relations (\ref{b-9}) and (\ref{b-10}) then lead to 
\bea
\langle \chi |\alpha (x) | \chi \rangle & = & \int \frac{dp}{2\pi g_s} u = \frac{1}{2\pi g_s} \beta (x) = \frac{1}{2\pi g_s} [P_{1+}(x)-P_{1-}(x)+P_{2+}(x)]~, \nonumber \\
\pi g_s \langle \chi|: \alpha (x)^2 : | \chi \rangle  & = & \int \frac{dp}{2\pi g_s} p~u = \frac{1}{2\pi g_s} [\frac{1}{2} \beta^2 (x) + w_1 (x)]=  \frac{1}{4\pi g_s} [P_{1+}(x)^2-P_{1-}(x)^2+P_{2+}(x)^2]\nonumber \\
\label{4-3}
\eea
Since the Dirac sea is infinite, one needs to subtract the ground state values to get finite quantities defined above. The relations (\ref{4-3}) show that 
\ben
w_1(x) = \langle \chi|: \alpha (x)^2 : | \chi \rangle - (\langle \chi |\alpha (x) | \chi \rangle)^2~.
\label{4-4}
\een
Therefore $w_1(x)$ is the quantum dispersion of the operator $\alpha (x)$ in this state. In the absence of a fold, i.e. a single intersection of a constant $x$ line with the fermi surface, the quantum dispersion vanishes. This is a coherent state of the chiral scalar. On the other hand, a nonzero $w_1(x)$ signifies a quantum dispersion which {\em which does not vanish in the classical limit}. An example of such a state is a state where a macroscopic number of fermions are excited into a band of states leaving a gap from the ground state fermi surface. Locally this is like a single fold. For details see \cite{fold1,fold2}.

\section{Chiral Bosonization formulae}
\label{appboson}

In this appendix we give some details of standard chiral bosonization formulae, following \cite{bosonization}. 
Consider a chiral fermion field $\psi (x)$ on a circle of cirumference $L$. The mode decomposition is given by
\ben
\psi (x) = \frac{1}{\sqrt{L}}\sum_{m=-\infty}^\infty\psi_m e^{\frac{2\pi m x}{L}}~, \qquad \psi^\dagger (x) = \frac{1}{\sqrt{L}}\sum_{m=-\infty}^\infty\psi^\dagger_m e^{-\frac{2\pi m x}{L}}~.
\label{b-1}
\een
The anticommutation relations  are
\ben
\{ \psi^\dagger_m,\psi_n \} = \delta_{mn} ~, \qquad \{ \psi_m,\psi_n \} = \{ \psi^\dagger_m,\psi^\dagger_n \}=0~.
\label{b-2}
\een
This leads to the standard anticommutators
\ben
\{ \psi (x) , \psi^\dagger (y) \} = \delta (x-y)
\label{b-3}
\een
with all other anticommutators vanishing.

Suppose we have a state $|0\rangle$ defined by
\ben
\psi_n |0\rangle = 0 \qquad {\rm for}~~n > 0~, \qquad
\psi^\dagger_n |0\rangle = 0 \qquad {\rm for}~~n \leq 0~.
\label{b-4}
\een
This leads to the normal ordering prescription
\bea
:\psi^\dagger_m \psi_n : & = & \psi^\dagger_m \psi_n \qquad n > 0~, \nonumber \\
:\psi^\dagger_m \psi_n : & = & -\psi_n\psi^\dagger_m \qquad n \leq 0~.
\label{b-5}
\eea
We can now define bosonic operators
\ben
\alpha_m = \sum_{n=-\infty}^\infty : \psi^\dagger_{-m+n}\psi_n : ~,
\label{b-6}
\een
which now obey
\ben
[ \alpha_m , \alpha_n ] = m \delta_{m+n,0}~.
\label{b-7}
\een
Normal ordering for these bononic operators is then given by
\bea
:\alpha_m \alpha_n: & = & \alpha_m \alpha_n \qquad n > 0~, \nonumber \\
:\alpha_m \alpha_n: & = & \alpha_n \alpha_m \qquad m > 0~.
\label{b-8}
\eea
This leads to a single chiral boson $\varphi (x)$
\ben
\alpha (x) \equiv \partial_x \varphi (x) = \frac{1}{L} \sum_{m=-\infty}^\infty \alpha_m e^{\frac{2\pi imx}{L}} = : \psi^\dagger (x) \psi (x) :~.
\label{b-9}
\een
The last equation follows from the mode relations (\ref{b-6}) and the normal orderings (\ref{b-5}). The fermion field can be expressed in terms of the chiral scalar as
\ben
\psi (x) = : e^{i\alpha (x)} :
\label{b-10}
\een 
and the current is
\ben
-\frac{i g_s}{2} : [ \psi^\dagger \partial_x \psi - (\partial_x \psi^\dagger) \psi ] : = \pi g_s :\alpha (x)^2: ~ . 
\label{b-11}
\een
In some state $|\chi\rangle$, from the definiton of the expectation value of the phase space density (\ref{0-7}), we therefore have the relations (\ref{4-4}).

%

\end{document}